\documentclass[journal]{vgtc}
\let\ifpdf\relax

\usepackage{mathptmx}
\usepackage{graphicx}
\usepackage{times}
\usepackage{amsmath,amscd}
\usepackage{amsfonts}
\usepackage[ruled,vlined]{algorithm2e}
\usepackage{enumerate}
\usepackage{multirow}
\usepackage{array}
\newtheorem{theorem}{Theorem}

\onlineid{000}
\vgtccategory{Research}
\vgtcinsertpkg

\title{Registration of Volumetric Prostate Scans using Curvature Flow}

\author{ Saad Nadeem, Rui Shi, Joseph Marino, Wei Zeng, Xianfeng Gu, and Arie Kaufman, \textit{Fellow, IEEE}}
\authorfooter{
\item
 Saad Nadeem, Rui Shi, Joseph Marino, Xianfeng Gu, and Arie Kaufman are with Stony Brook University, Stony Brook, NY 11794-4400.\\E-mail: \{sanadeem, ruishi, jmarino, gu, ari\}@cs.stonybrook.edu.
\item
 Wei Zeng is with Florida International University.\\Email: wzeng@cs.fiu.edu
}

\abstract{
 Radiological imaging of the prostate is becoming more popular among researchers and clinicians in searching for diseases, primarily cancer. Scans might be acquired with different equipment or at different times for prognosis monitoring, with patient movement between scans, resulting in multiple datasets that need to be registered. For these cases, we introduce a method for volumetric registration using curvature flow. Multiple prostate datasets are mapped to canonical solid spheres, which are in turn aligned and registered through the use of identified landmarks on or within the gland. Theoretical proof and experimental results show that our method produces homeomorphisms with feature constraints. We provide thorough validation of our method by registering prostate scans of the same patient in different orientations, from different days and using different modes of MRI. Our method also provides the foundation for a general group-wise registration using a standard reference, defined on the complex plane, for any input. In the present context, this can be used for registering as many scans as needed for a single patient or different patients on the basis of age, weight or even malignant and non-malignant attributes to study the differences in general population. Though we present this technique with a specific application to the prostate, it is generally applicable for volumetric registration problems.
} 


\CCScatlist{ 
\CCScat{Computer Graphics}{I.3.3}{Data Registration}{Geometry-based Techniques, Medical Visualization, Mathematical Foundations for Visualization}
}

\teaser{
\centering
\includegraphics[scale=0.85]{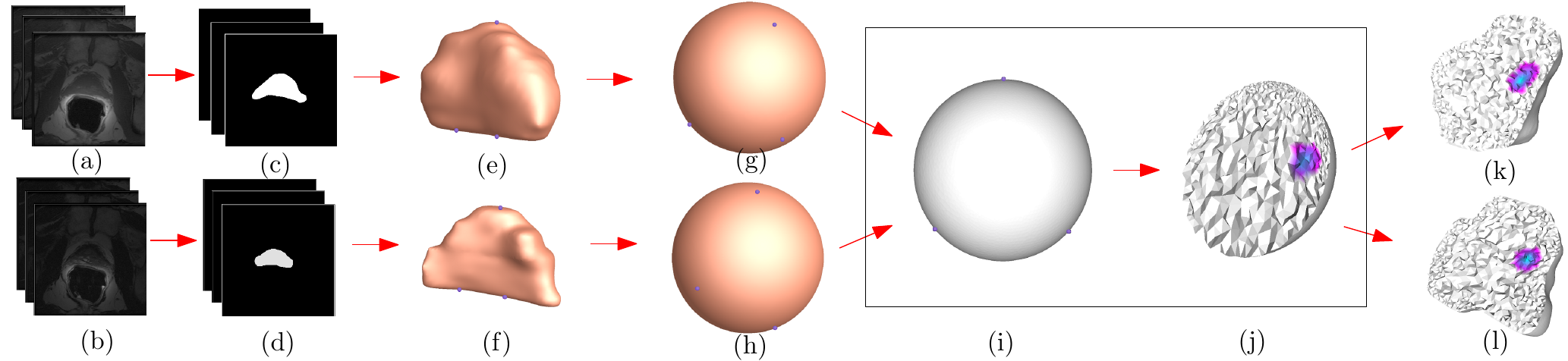}
\caption{Registration of $T_1$ and $T_2$ axial MR scans for the same patient on the same day. MR scans are taken from (a) $T_1$ and (b) $T_2$ axial view and segmented (c \& d). These segments are reconstructed into a surface (e \& f) and mapped onto respective spheres and tetrahedralized (g \& h). These spheres are then registered using surface feature points onto another sphere (i). On this registered sphere, we can mark regions (j) and since our mapping is diffeomorphic, these regions are mapped back to the original tetrahedralized prostate shapes (k \& l).}
\label{fig:teaser}
}

\renewcommand{\manuscriptnotetxt}{}

\nocopyrightspace

\begin{document}


\firstsection{Introduction}

\maketitle

Cancer of the prostate is the second leading cause of cancer-related
mortality among males in the United States, and is the most commonly
diagnosed cancer in general~\cite{seer:2005}.  Although it is such
a common cancer, diagnosis methods remain primitive and inexact.
Detection relies primarily on the use of a simple blood test to
check the level of prostate specific antigen (PSA) and on the digital
rectal examination (DRE).  If an elevated PSA level is found, or if
a physical abnormality is felt by the physician during a DRE, then
biopsies will be performed.  Though guided by transrectal ultrasound
(TRUS), these biopsies are inexact, and large numbers (143 in one
case) are often necessary to try and retrieve a sample from a
cancerous area~\cite{loch:2007:wju}.  More recently, it has been
noted that Magnetic Resonance Imaging (MRI) can be used for the
detection of prostate cancer~\cite{hersh-2004-cc}.

Multiple MR images obtained with different settings are necessary
for the detection of prostate cancer.  Most commonly used is a
combination of $T_{1}$ and $T_{2}$ image sequences.  $T_{2}$ images
are generally of higher quality than $T_{1}$ images, and prostate
cancer will appear as areas of reduced intensity.  The acquisition of these MR image sequences is often done with varying
orientations and resolutions per sequence.  In cases where the image
sequences are acquired during a single session without the patient
moving, the resulting volumes will be naturally registered in world
space.  However, there are many scenarios, where registration methods are
necessary for multiple sequences.  One such reason would be the
situation where a patient shifts during a session, such that the
patient's body position is not constant, and thus the volumes will
not orient themselves properly with respect to each other in 3D
space.  Registration is also required when the sequences are
acquired at different times.  This could be a common occurrence to
monitor the progression of the tumor(s) during watchful waiting or
brachytherapy.  Thirdly, it is possible that one might want to
correlate the MR images to histology information in order to confirm
results during development and testing of a system.

In this paper, we proposed a novel volumetric registration framework,
which can reduce the work of radiologists, and help to register the datasets, to monitor the progression of abnormalities,
and to facilitate the diagnoses of prostate diseases. The pipeline consists of reconstruction, mapping, and registration. We first reconstruct the volume of the prostate from $T_1$ or $T_2$ MR images. Then we utilize our volumetric mapping algorithm to map the prostate to a solid ball. Finally we register the solid balls obtained from the mapping, resulting in a registration of the original prostates.

Our volumetric registration framework is based on a volumetric
parameterization algorithm using discrete volumetric Ricci flow, by which a ball-shape volume is mapped to a canonical domain, i.e., a unit ball. This parameterization technique is a homeomorphism, which means one-to-one and continuous. This property is important to medical imaging and visualization because we do not want to lose the fidelity of the data, or miss any part which could possibly be an abnormality, or a sign of the diseases.
This mapping also has a low stretching energy in a physical sense, so that the shape distortion is limited for small regions after the mapping.
After mapping two prostates to the unit ball, we use anatomic feature points to further align and register them. We have thoroughly tested our registration method on data from ACRIN 6659 study, Prostate MR Image Database \cite{MRIDatabase:2008} and Stony Brook UHMC Database for scans acquired in different orientations, different modes of MRI and different days. The results show that we can easily handle all these cases and give good registration.

The main contributions of this work are as follows:
\begin{enumerate}
  \item We propose a novel volumetric registration framework
  \item We show registration of prostate MR scans in different orientations, different modes and different days (for prognosis monitoring)
\end{enumerate}

The rest of the paper is organized as follows. In Section 2 we review the previous work on volume registration. Then we introduce the theoretical foundation of
our algorithm in Section 3, followed by the details of our framework in Section 4. Evaluation methods and experimental results are shown in Section 5 and 6, respectively. Finally, we conclude our discussion and sketch the future research plan in Section 7.

\section{Related Works}

There have been some related works on medical volumetric registration, but most of them are $2D$ based, which cannot ensure the accuracy, or even the one-to-one property. In particular, registration of prostate volumes from MRI has often relied on the
use of a similarity matrix and the application of a deformation
field to the original MR image slices~\cite{bharatha:2001:medphys}.
More recently, it has been suggested to perform the nonrigid
registration by extracting tetrahedral grids from the volume
datasets and deforming these grids by using the finite element
method~\cite{daische:2004:miccai}. We also use the tetrahedral grid
structure, but perform the deformation through the use of the volumetric curvature flow algorithm.

Ricci flow \cite{ric88} is a powerful tool for geometric analysis.
It has been applied to prove the Poincar\'e conjecture successfully.
Surface Ricci flow presents an efficient approach to computing the uniformization theorem \cite{Forster91}. \emph{Ricci flow} refers to the process of deforming Riemannian metric $\mathbf{g}$ proportional to the curvature, such that the curvature $K$ evolves according to a heat diffusion process, eventually the curvature becomes constant everywhere \cite{Chow06}. Discrete surface Ricci flow \cite{jin08,yongliang} generalizes the curvature flow method from smooth
surface to discrete triangular meshes. The key insight to discrete Ricci flow is based on the following observation: conformal mappings
transform infinitesimal circle fields to infinitesimal circle fields. Furthermore, Ricci flow can be used to construct the unique conformal Riemannian metric with prescribed Gaussian curvatures. The resulting conformal surface mappings are free of angle distortion.

Volumetric mapping is a generalization of the surface mapping problem.
In theory, three dimensional manifolds (volumetric) also admit uniformization metrics, which induce constant sectional curvature.
The only existing approach for generalizing of surface Ricci curvature flow to volume was presented by Yin et al. \cite{ISVC_HYPERBOLIC_08}.
It generalizes the discrete curvature flow from surfaces to \emph{hyperbolic} 3-manifolds with complete geodesic boundaries.
The metric deforms according to the curvature, until the curvature is constant everywhere. However, it can only support the hyperbolic 3-manifolds and cannot handle the cases of topological ball. In this paper, we present the discrete volumetric Ricci flow to compute the volumetric mappings for topological balls, such as the volumetric prostates. To the best of our knowledge, the proposed method is the first work to generalize the discrete surface curvature flow to 3-manifolds with solid ball topology.

The uniformization metrics on 3-manifolds have great potential for volumetric parameterization, volumetric shape matching and registration, and volumetric shape analysis.
In this work, we propose to use this discrete volumetric curvature flow for registration of topological ball volumes with large deformations. Compared to the previous methods, our proposed method can generate \emph{diffeomorphisms} (one-to-one and onto mappings). It is \emph{global and steady}, and can handle \emph{large deformations}.

There are some related works on volumetric mapping. The volumetric harmonic map depends on volumetric Laplacian; Zhou et al. \cite{ss6} have applied volumetric
graph Laplacian to large mesh deformation. Harmonicity in volumes can be similarly defined via the vanishing Laplacian, which
governs the smoothness of the mapping function. Wang et al. \cite{ss7} have studied the formula of
harmonic energy defined on tetrahedral meshes and computed the discrete volumetric
harmonic maps by a variational procedure. Volumetric parameterization using fundamental
solution method \cite{ss8} was applied to volumetric deformation and
morphing. Other than that, harmonic volumetric parameterization for cylinder volumes
is applied for constructing tri-variate spline fitting ~\cite{ss9}. All the above approaches rely
on volumetric harmonic maps. Unfortunately, these volumetric harmonic maps cannot guarantee bijective mappings even though the
target domain is convex. Besides the volumetric harmonic map, there is another approach of mean value coordinates for closed triangular meshes \cite{ss10, ss11}. Mean value coordinates are a powerful and flexible tool to define a map between two volumes. However, there is no guarantee
that the computed map is a diffeomorphism.

Our approach differs intrinsically from these existing approaches in two ways. First,
we solve the curvature flow for \emph{arbitrary topological balls} (prostate volumes are of this case).
 As a result, the curvature flow induced map is guaranteed to be a diffeomorphism.
 Second, we use the theory of \emph{Ricci flow} rather
than the conventional volumetric harmonic map~\cite{ss7}.
\section{Theory}

This section briefly introduces the theoretic foundation of our discrete curvature flow method.

\subsection{Surface Ricci Flow}

Let $S$ be a smooth surface with a Riemannian metric $\mathbf{g}$.
One can always find isothermal coordinates $(u,v)$ for $\mathbf{g}$
locally, which satisfies
\begin{equation}
    \mathbf{g} = e^{2\lambda(u,v)} (du^2 + dv^2).
\end{equation}
where $\lambda$ is a conformal factor.
The Gaussian curvature of the surface is given by
\[
    K(u,v) = -\Delta_\mathbf{g} \lambda,
    \label{eqn:gaussian_curvature}
\]
where $\Delta_\mathbf{g} =
e^{-2\lambda(u,v)}(\frac{\partial^2}{\partial u^2} +
\frac{\partial^2}{\partial v^2})$ is the Laplace-Beltrami operator
induced by $\mathbf{g}$. Although the Gaussian curvature is
intrinsic to the Riemannian metric, the total Gaussian curvature is
a topological invariant: the total Gaussian curvature of a closed
metric surface is
\[
    \int_S K dA = 2\pi \chi(S),
\]
where $\chi(S)$ is the Euler number of the surface. Suppose
$\mathbf{g}_1$ and $\mathbf{g}_2$ are two Riemannian  metrics on the
smooth surface $S$. If there is a differential function $\lambda : S
\to \mathbb{R}$, such that
\[
    \mathbf{g}_2 =  e^{2\lambda} \mathbf{g}_1,
\]
then the two metrics are \emph{conformal equivalent}. Let the
Gaussian curvatures of $\mathbf{g}_1$ and $\mathbf{g}_2$ be $K_1$
and $K_2$ respectively. Then
 they satisfy the following \emph{Yamabe equation}
\begin{equation}
    K_2 = \frac{1}{e^{2\lambda}}( K_1 - \Delta_{\mathbf{g}_1} \lambda ).
\end{equation}

Suppose the metric $\mathbf{g}=(g_{ij})$ in local coordinate.
Hamilton introduced the Ricci flow as
\begin{equation}
    \frac{d}{dt}g_{ij}(t) = -K(t) g_{ij}(t).
\end{equation}
During the flow, the Gaussian curvature will evolve according to a
heat diffusion process.
\begin{theorem}[Hamilton and Chow] \cite{Chow06}
Suppose $S$ is a closed surface with a Riemannian metric. If the
total area is preserved, the surface Ricci flow will
 converge to a Riemannian metric of constant  Gaussian curvature.
\end{theorem}

\subsection{General Ricci Flow}

This section uses the Einstein summation convention. Given
coordinate functions $x^i, i=0,1,2$, the tangent vector can be
described by its components in the basis $\mathbf{e}_i =
\frac{\partial}{\partial x^i}$. The Riemannian metric tensor is
given by
\[
    g_{ij} = \langle \mathbf{e}_i,\mathbf{e}_j \rangle.
\]
Let vector fields $\mathbf{v}=v^i\mathbf{e}_i$ and
$\mathbf{u}=u^j\mathbf{e}_j$. The Riemannian metric defines an inner
product
\[
    \langle \mathbf{u},\mathbf{v} \rangle = g_{ij}v^iu^j.
\]
The so-called Christoffel symbols are given by
\begin{equation}
    \Gamma_{ij}^k = \frac{1}{2} g^{kr}(\frac{\partial g_{rj}}{\partial x^i}+ \frac{\partial g_{ri}}{\partial x^j}-\frac{\partial g_{ij}}{\partial x^r}),
\end{equation}
where $(g^{kr})$ is the inverse matrix of $(g_{ij})$.

The covariant derivative of a basis vector along a basis vector is a
vector, and is given by
\[
    \nabla_{\mathbf{e}_i}\mathbf{e}_j = \Gamma_{ij}^k \mathbf{e}_k.
\]
We get
\[
    \nabla_\mathbf{v} \mathbf{u} = (v^iu^j\Gamma_{ij}^k + v^i \frac{\partial u^k}{\partial
    x^i})\mathbf{e}_k.
\]
The curvature tensor of the manifold is given by
\[
    R(\mathbf{u},\mathbf{v})\mathbf{w} = \nabla_\mathbf{u}\nabla_\mathbf{v} \mathbf{w} - \nabla_\mathbf{v}\nabla_\mathbf{u} \mathbf{w} - \nabla_{[\mathbf{u},\mathbf{v}]}\mathbf{w}.
\]
Curvature tensor measures non-commutativity of the covariant
derivative. The \emph{sectional curvature} is given by
\begin{equation}
    K(\mathbf{u},\mathbf{v}) = \frac{<R(\mathbf{u},\mathbf{v})\mathbf{v},\mathbf{u}>}{|\mathbf{u} \wedge \mathbf{v}|^2}.
\end{equation}
where $\wedge$ is the wedge product.
Intuitively, vector $u$ and $v$ span a tangent plane $\sigma$ in
the tangent space at a point $p\in M$. All the geodesics emanating
from $p$ and tangent to the plane $\sigma$ form a surface $S\subset
M$. The Gaussian curvature of $S$ at $p$ is $K(u,v)$.

Assume $\{e_i\}$ is an orthonormal basis of the tangent space at
$p$, the \emph{scalar curvature} is the trace of the curvature
tensor,
\[
Sc(p)=\sum_{ij}<R(\mathbf{e}_i,\mathbf{e}_j)\mathbf{e}_j,\mathbf{e}_i>.
\]
\emph{Ricci curvature} is a linear operator on tangent space at a
point,
\[
    Ric(\mathbf{u})=\sum_i R(\mathbf{u},\mathbf{e}_i)\mathbf{e}_i.
\]
For general Riemannian manifold, Ricci flow is given by
\begin{equation}
    \frac{d}{dt} g_{ij}(t) = -2(R_{ij}(t)-\frac{2\pi\chi(s)}{A(0)}).
\end{equation}

The evolution behavior of Ricci flow on 3-manifolds is much more
complicated, because singularities will emerge in the flow.
We refer readers to \cite{Chow06} for detailed explanation.

\subsection{Discrete Surface Ricci Flow}
\label{sec:srf}
Discrete surface Ricci flow generalizes the curvature flow method
from smooth surface to discrete triangular meshes. The key insight
to discrete Ricci flow is based on the following observation:
conformal mappings transform infinitesimal circle fields to
infinitesimal circle fields. Discrete Ricci flow replaces
infinitesimal circles by circles with finite radii, and modifies the
circle radii to deform the discrete metric, to achieve the desired
curvature.

In engineering fields, surfaces are approximated by their
triangulations, the triangle meshes.  A triangle mesh $\Sigma$ is a
$2$ dimensional simplical complex embedded in $\mathbb{R}^3$.

The discrete Riemannian metric is defined as the set of edge
lengths,
\[
l: E \to \mathbb{R}^+, \label{eqn:discrete_metric}
\]
as long as, for each face $[v_i,v_j,v_k]$, the edge lengths satisfy
the triangle inequality: $l_{ij} + l_{jk} > l_{ki}$. The edge
lengths determine the corner angles by cosine law. The discrete
Gaussian curvature $K_i$ at a vertex $v_i\in \Sigma$ can be computed
as the angle deficit,
\begin{equation}
    K_{i}= \left\{
    \begin{array}{rl}
    2\pi - \sum_{[v_i,v_j,v_k]\in \Sigma} \theta_i^{jk},
    & v_i \not\in \partial \Sigma\\
    \pi - \sum_{[v_i,v_j,v_k]\in \Sigma} \theta_i^{jk},
    & v_i \in \partial \Sigma
    \end{array}
    \right.
\label{eqn:discrete_gaussian_curvature}
\end{equation}
where $\theta_i^{jk}$ represents the corner angle attached to vertex
$v_i$ in the face $[v_i,v_j,v_k]$, and $\partial \Sigma$ represents
the boundary of the mesh.

The Gauss-Bonnet theorem  holds on meshes as follows.
\begin{equation}
\sum_{v_i \in V} K_i  = 2\pi \chi(M).
\label{eqn:discrete_gauss_bonnet}
\end{equation}
where M is a mesh.

In the smooth case, \emph{Conformal deformation of a Riemannian
metric} is defined as
\[
    \mathbf{g} \to e^{2\lambda} \mathbf{g}, \quad \lambda : S \to \mathbb{R}.
\]
In the discrete case, there are many ways to define conformal metric
deformation. Generally, we associate each vertex $v_i$ with a circle
$(v_i,\gamma_i)$ centered at $v_i$ with radius $\gamma_i$.

\begin{figure}
 \centering
\begin{tabular}{cc}
\includegraphics[height=3.6cm]{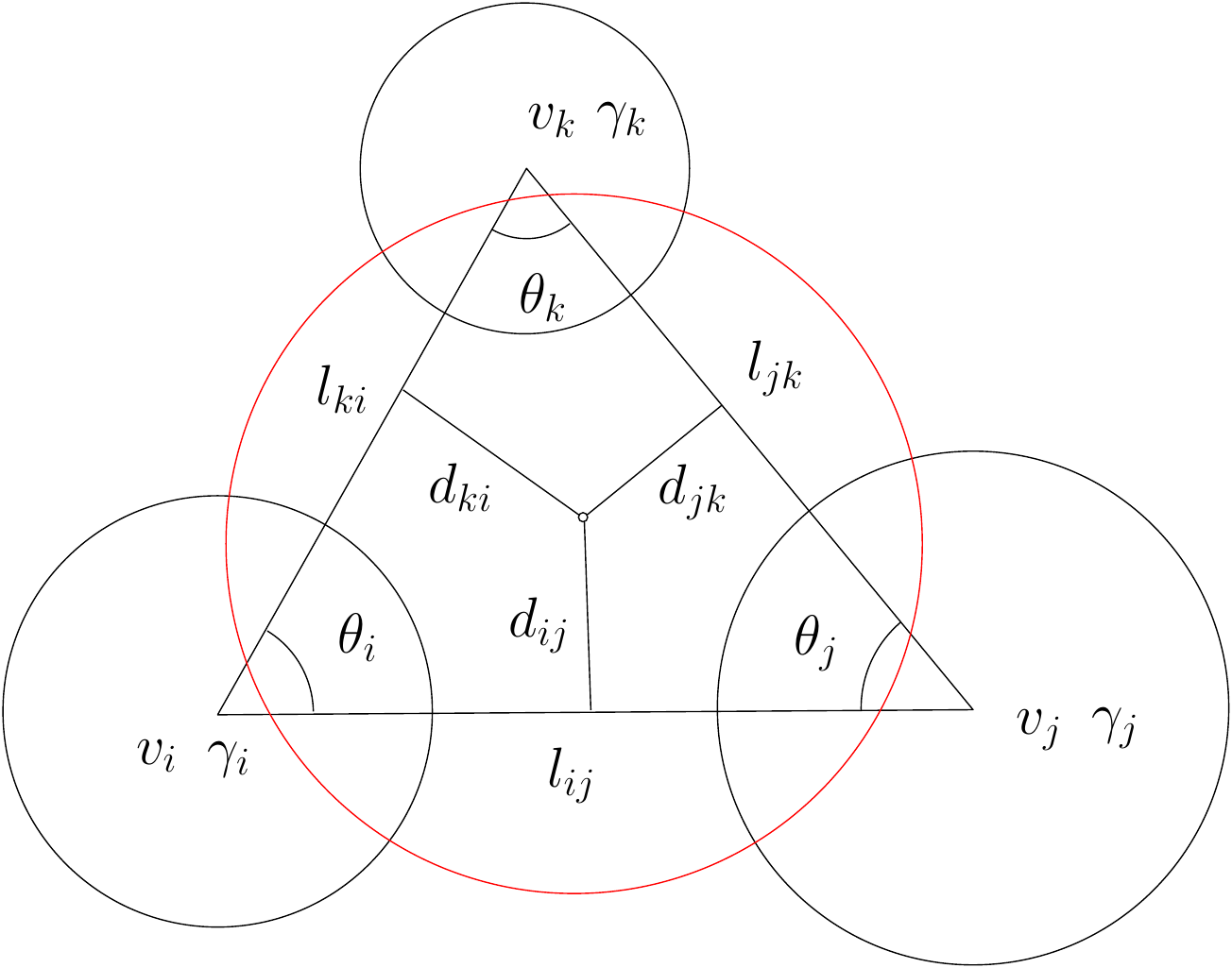} &
\includegraphics[height=3.0cm]{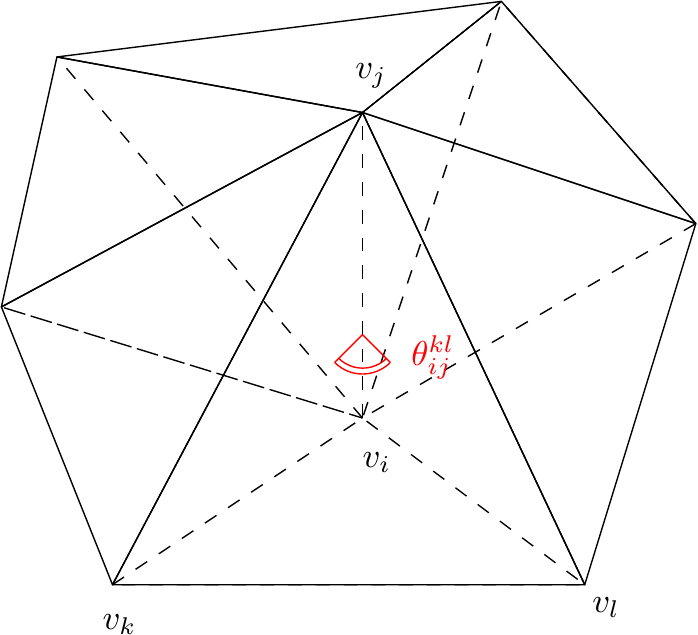} \\
(a) & (b)
\end{tabular}
\caption{Configuration of discrete Ricci flow (a) Surface case, (b) Volume case. (a) Inversive circle packing. (b) Discrete edge Ricci Curvature.
 \label{fig:circle}}
\end{figure}

On an edge $[v_i,v_j]$, two circles are separated, the edge length
is given by,
\begin{equation}
    l_{ij}^2 = \gamma_i^2 + \gamma_j^2 + 2\gamma_i \gamma_j I_{ij},
\end{equation}
where $I_{ij}$ is the so-called \emph{inversive distance}. During
the conformal deformation, the radii of circles can be modified, but
the inversive distances are preserved. There exists a unique circle,
the so called \emph{radial circle}, that is orthogonal to three
vertex circles. The radial circle center is denoted as $o$ (see Figure~\ref{fig:circle}(a)).

\begin{figure}
 \centering
\begin{tabular}{cc}
\includegraphics[height=4.0cm]{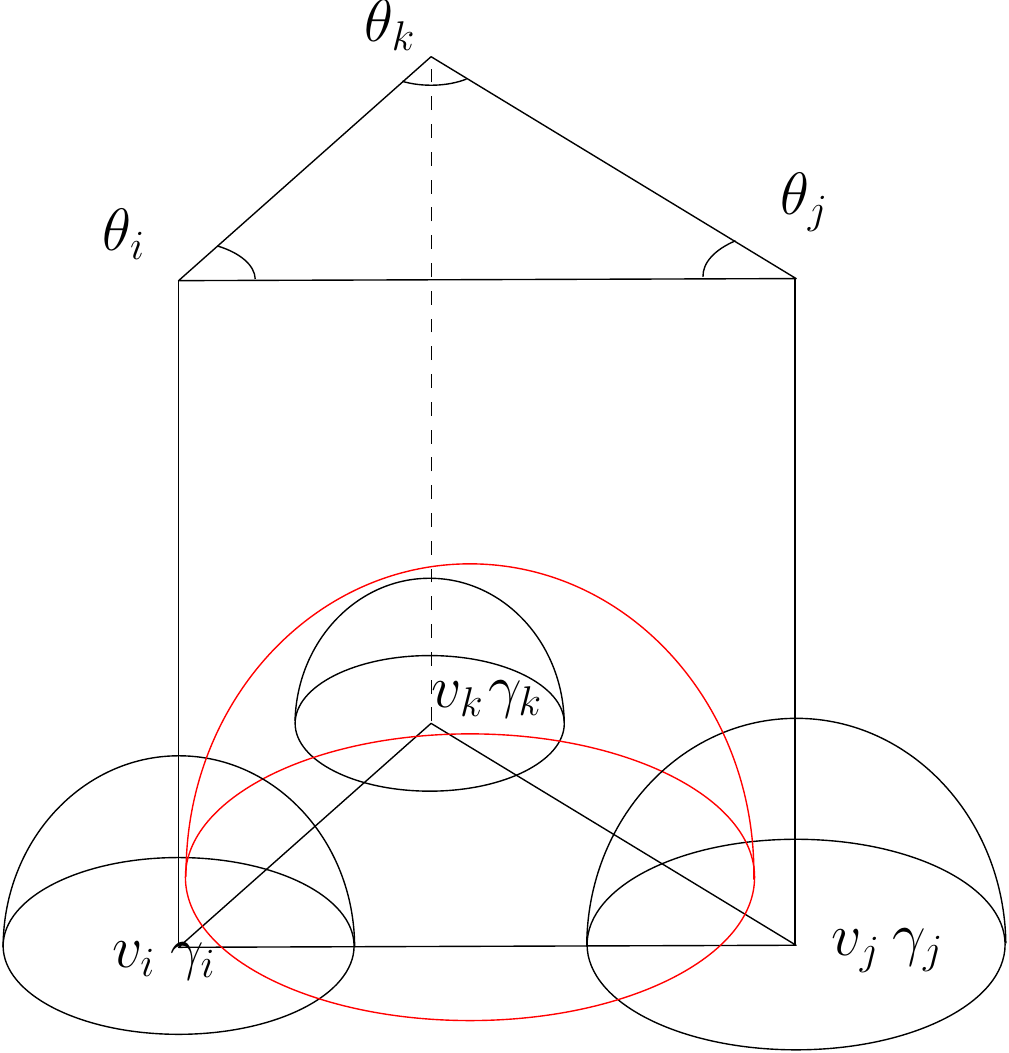} &
\includegraphics[height=4.0cm]{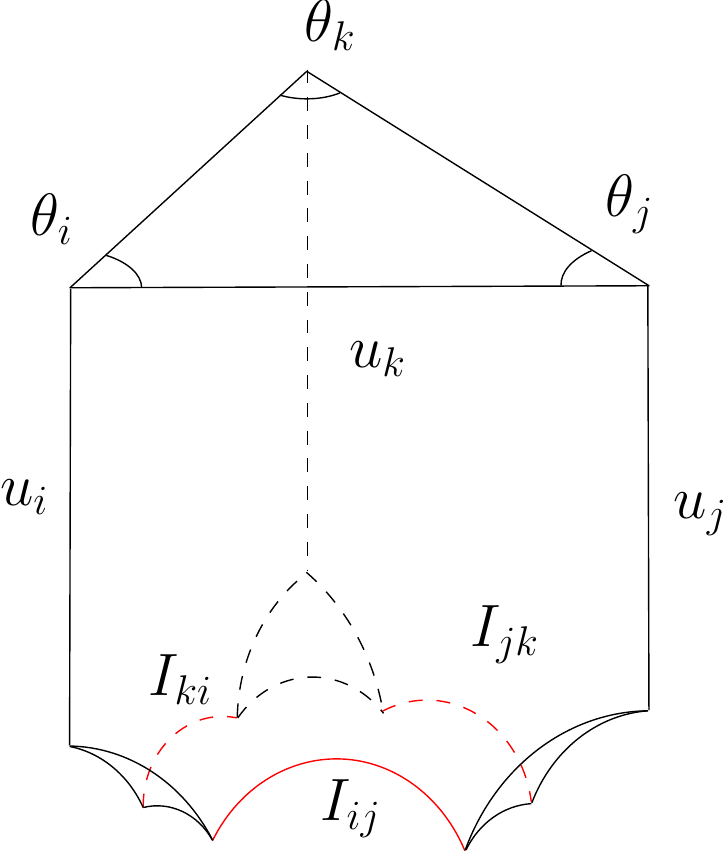} \\
(a) & (b)
\end{tabular}
\caption{Geometric interpretation of discrete surface Ricci energy. The inversive distance circle packing on each face (a) truncates a hyperbolic tetrahedron (b). The Ricci energy is the total hyperbolic volumes of the truncated tetrahedra.
 \label{fig:energy}}
\end{figure}

Let $u_i$ be the logarithm of $\gamma_i$,  the \emph{discrete Ricci
flow} is defined as follows:
\begin{equation}
\frac{du_i(t)}{dt} = (\bar{K}_i-K_i ),\\
\label{eqn:discrete_ricci_flow}
\end{equation}
where $\bar{K}_i$ is the user defined target curvature and $K_i$ is
the curvature induced by the current metric. The discrete Ricci flow
has exactly the same form as the smooth Ricci flow, which
conformally deforms the discrete metric according to the Gaussian
curvature.

The discrete Ricci flow can be formulated in the variational
setting, namely, it is a negative gradient flow of a special energy
form, the so-called \emph{entropy energy}. The energy is given by
\begin{equation}
f(\mathbf{u})=\int_{\mathbf{u_0}}^\mathbf{u} \sum_{i=1}^n
(\bar{K_i}-K_i) du_i, \label{eqn:discrete_ricci_energy}
\end{equation}
where $\mathbf{u_0}$ is an arbitrary initial metric. Figure \ref{fig:energy} shows the geometric interpretation of the discrete Ricci energy. Each triangle with a circle pattern corresponds to a truncated hyperbolic tetrahedron. As shown in frame (a), the triangle is laid on the xy-plane, a prism is built based on the triangle. Through each circle, one hemisphere is constructed. The intersection between the inside of the prism and the outside of all hemispheres is a truncated tetrahedron. The upper half space $\{z>0\}$ is treated as three dimensional hyperbolic space $\mathbb{H}^3$ with the Riemannian metric
\[
    \frac{dx^2+dy^2+dz^2}{z}.
\]
The truncated tetrahedron is treated as a hyperbolic tetrahedron, with edge lengths $\{u_i, u_j, u_k, I_{ij}, I_{jk}, I_{ki}\}$. The dihedral angles on edges $\{u_i,u_j,u_k\}$ are $\{\theta_i, \theta_j,\theta_k\}$ (see Figure \ref{fig:energy}(b)). Denote the hyperbolic volume of the tetrahedron as $V$, then according to Schlafli formula:
\begin{equation}
    \frac{dV}{d\theta_i } = u_i.
\end{equation}
Then the Ricci energy is the Legendre dual to the volume,
\begin{equation}
   E(v_i,v_j,v_k)= u_i\theta_i + u_j \theta_j + u_k \theta_k - V([v_i,v_j,v_k]),
\end{equation}
therefore $dE = \theta_i du_i + \theta_j du_j + \theta_k du_k$. The
Ricci energy for the whole mesh is the summation of those of all
faces,
\begin{equation}
    E(M) = \sum_{i,j,k} E( [v_i,v_j,v_k]).
\end{equation}

Computing the desired metric with user-defined curvature
$\{\bar{K}_i\}$ is equivalent to minimizing the discrete entropy
energy. Because the discrete entropy energy is locally convex, it
has global minima.

The Hessian matrices for discrete entropy are positive definite. The
energy can be optimized using Newton's method. The Hessian matrix
can be computed using the following formula. Given a triangle
$[v_i,v_j,v_k]$, let $d_{ij}$ be the distance from the radial circle
center to the edge $[v_i,v_j]$ (see Figure \ref{fig:circle}(a)), then
\[
    \frac{\partial \theta_i}{\partial u_j} = \frac{d_{ij}}{l_{ij}},
\]
furthermore
\[
    \frac{\partial \theta_j}{\partial u_i} = \frac{\partial \theta_i}{\partial u_j},\quad
    \frac{\partial \theta_i}{\partial u_i} = -\frac{\partial \theta_i}{\partial u_j}-\frac{\partial \theta_i}{\partial u_k}.
\]
We define the edge weight $w_{ij}$ for edge $[v_i,v_j]$, which is
adjacent to $[v_i,v_j,v_k]$ and $[v_j,v_i,v_l]$ as
\begin{equation}
    w_{ij} = \frac{d_{ij}^k+d_{ij}^l}{l_{ij}}.
\end{equation}
The Hessian matrix $H=(h_{ij})$ is given by the discrete Laplace
form
\begin{equation}
h_{ij} = \left\{
\begin{array}{rl}
0,&   [v_i,v_j]\not\in E\\
-w_{ij}, & i \neq j\\
\sum_{k} w_{ik}, & i = j
\end{array}
\right.
\end{equation}
Algorithmic details for discrete surface Ricci flow can be found in
\cite{jin08} and \cite{yongliang}.

\subsection{Discrete Volumetric Ricci Flow}

Volumetric data are represented as tetrahedral meshes. Suppose
$[v_i,v_j,v_k,v_l]$ is a tetrahedron, the  dihedral angle on edge
$e_{ij}=[v_i,v_j]$ is denoted as $\theta_{ij}^{kl}$ (see Figure \ref{fig:circle}(b)). If the edge is
an interior edge, (i.e. $e_{ij}$ is not on the boundary surface),
the discrete Ricci curvature is defined as
\begin{equation}
    K(e_{ij}) = 2\pi - \sum_{kl} \theta_{ij}^{kl}.
\end{equation}
If $e_{ij}$ is on the boundary surface, its curvature is given by
\begin{equation}
    K(e_{ij}) = \pi - \sum_{kl} \theta_{ij}^{kl}.
\end{equation}
The discrete curvature flow for volume is defined by
\begin{equation}
    \frac{d}{dt} l_{ij} = K_{ij},
\end{equation}
where $l_{ij}$ is the edge length of $e_{ij}$.

Similar to the discrete surface Ricci flow, volumetric curvature
flow is also variational, which is the gradient flow of the
following energy:
\begin{equation}
E([v_i,v_j,v_k,v_l]) = Vol([v_i,v_j,v_k,v_l])-\sum_{ij}
l_{ij}\theta_{ij}^{kl},
\end{equation}
where $Vol$ is the volume of the tetrahedron. Therefore, the total
energy for the whole tetrahedron mesh is the summation of those of
each tetrahedron,
\begin{equation}
E(M) = \sum_{i,j,k,l} E([v_i,v_j,v_k,v_l]).
\end{equation}

In the procedure of the curvature flow, the image of each tetrahedra is non-degenerated. Therefore, the mapping is a homeomorphism.

Unlike surface case, the volumetric Ricci energy is not convex. The
const curvature metric corresponds to a critical point of the total
energy, which is a saddle point. Therefore, in the computational
process, the choice of initial Riemannian metric is important. In
practice, we use homotopy method explained as follows.

In our current project, we deform the prostate volume $M$ to the
unit solid ball $\mathbb{D}^3$. The initial metric on $M$ is denoted
as $\mathbf{g}_0$. The boundary surface of $M$ is denoted as
$\partial M$. First we use discrete surface Ricci flow method to map
$\partial M$ to the unit sphere $\mathbb{S}^2$. This gives us the
final metric on the boundary surface $\partial M$, denoted as
$\mathbf{g}_1$. Then we define a one parameter family of Riemannian
metrics on $\partial M$ as
\begin{equation}
    \mathbf{g}(t) = (1-t) \mathbf{g}_0 + t \mathbf{g}_1, \quad t \in
    [0,1].
\end{equation}

For each $t$, we fix the metric on the boundary, and then use volumetric
curvature flow to compute a metric, such that all interior edge
curvatures are zeros. Then we use $\mathbf{g}(t)$ as the initial
metric, to compute $\mathbf{g}(t+\delta t)$. This homotopy method
greatly improves the stability and robustness of volumetric
curvature flow.

\section{Algorithm}

Our volume registration pipeline consists of three steps: reconstruction, parameterization and registration. First the tetrahedral mesh is reconstructed after we segment the prostate from the scans. Then each scan is parameterized to a canonical unit ball respectively. After that, the resulting balls are registered using marked feature points, which could give us a registration of the prostates. Here we explain in details the algorithms in each step of our pipeline.

\subsection{Volume Reconstruction}

The segmentation of the prostate from the surrounding anatomy is
the first step. Though there has been research into the
automatic segmentation of the prostate from axial MR
images~\cite{zwiggelaar:2003:ibpria}, it is not general to all
possible types of images. Since our focus is on the registration
of the prostate, segmentation is outside the scope of this paper, and therefore, we apply manual segmentation to the datasets. After we extract the prostate from the raw image slices, we apply the Marching Cubes algorithm~\cite{rmarching} to build the corresponding boundary surface $S$ of the prostate, where tricubic interpolation~\cite{rtricubic} is used for resampling in case of possible different resolutions in the three dimensions.

The triangle mesh which we get from the previous step acts as the boundary constraint of the volume tetrahedralization. Then we use Tetgen~\cite{ss16} to generate a tetrahedral mesh for a given surface mesh $S$ to meet the boundary constraints. To improve the meshing quality,
we employ the variational tetrahedral meshing technique~\cite{ss4} which can significantly reduce
the slivers and produce well-shaped tetrahedral meshes, as shown in Figure~\ref{fig:tetMesh}(a).

\subsection{Volumetric Parameterization}

Here we assume the volume is represented as a tetrahedral mesh $M=(V,E,F,T)$ as follows. $V$ is the vertex set; $E$ is the edge set; $F$ is the face set; and $T$ is the tetrahedron set. Suppose $t_{ijkl}$ is a tetrahedron with vertices $\{v_i,v_j ,v_k,v_l\}$, and $e_{ij}$ is the edge connecting vertices $v_i$ and $v_j$.

\paragraph{Surface Parameterization}
We first use our surface curvature flow method (Algorithm \ref{alg:a1}) to calculate a parameterization $f:\partial M\to\mathbb{S}^2$ from the prostate boundary surface to the unit sphere.

\begin{algorithm}[!]
\caption{{Surface parameterization of a sphere.}}
\SetKwInOut{Require}{Require}
\Require{The prostate volume $M$}
\begin{enumerate}
\item Extract the boundary surface $\partial M$ of the prostate
volume $M$.
\item Remove one triangle $f_0=[v_0,v_1,v_2]$ from $\tau: \partial M$,
isometrically map the triangle onto the plane, denote the image
triangle on the plane as $[\tau(v_0),\tau(v_1), \tau(v_2)]$
\item Using discrete surface Ricci flow introduced in Section \ref{sec:srf} to compute a conformal map from $\partial M - f_0 \to
\mathbb{R}^2$ by setting the target curvature
\[
    \bar{K}_i = 0, \quad \forall i > 2.
\]
 \item Fill in the removed face $\tau(f_0)$. Scale the image of $\tau(\partial M)$.
 \item Use stereo-graphic projection to map $\tau(\partial M)$ to
 the unit sphere. The north pole is inside the image of $f_0$.
\end{enumerate}
\normalsize
\label{alg:a1}
\end{algorithm}

\paragraph{Metric Homotopy Method for Volumetric Curvature Flow}

After we get the boundary surface mapping, we can compute the
discrete volume Ricci flow using the surface mapping as boundary
constraint. In order to improve the stability and the robustness of
the volumetric curvature flow, we use metric homotopy method to map
the the prostate volume to the solid unit ball, the mapping is
denoted as $F:M\to\mathbb{D}^3$.

\begin{figure}[tb]
    \centering
 \begin{tabular}{cc}
 \includegraphics[width=1.57in]{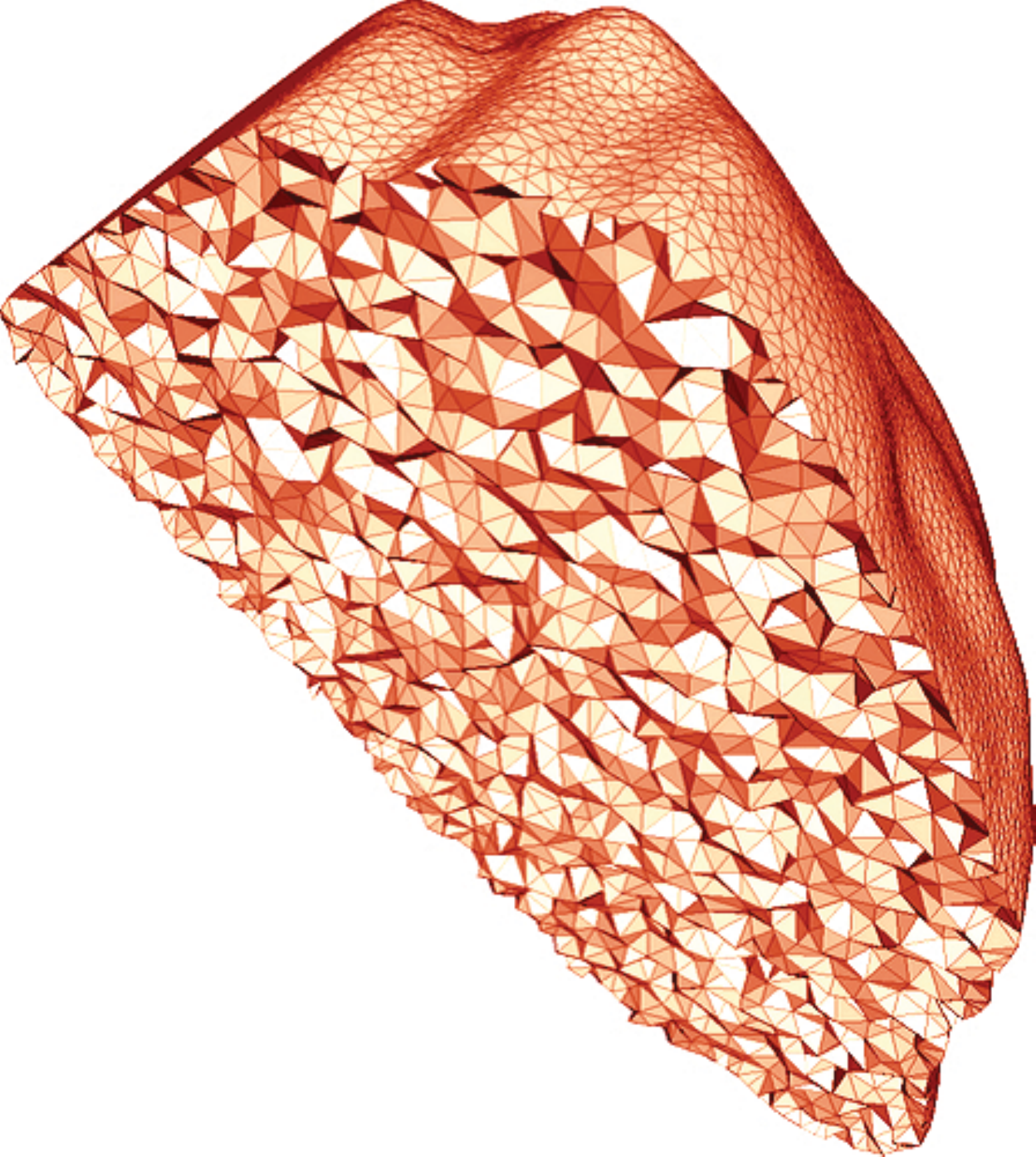} &
 \includegraphics[width=1.57in]{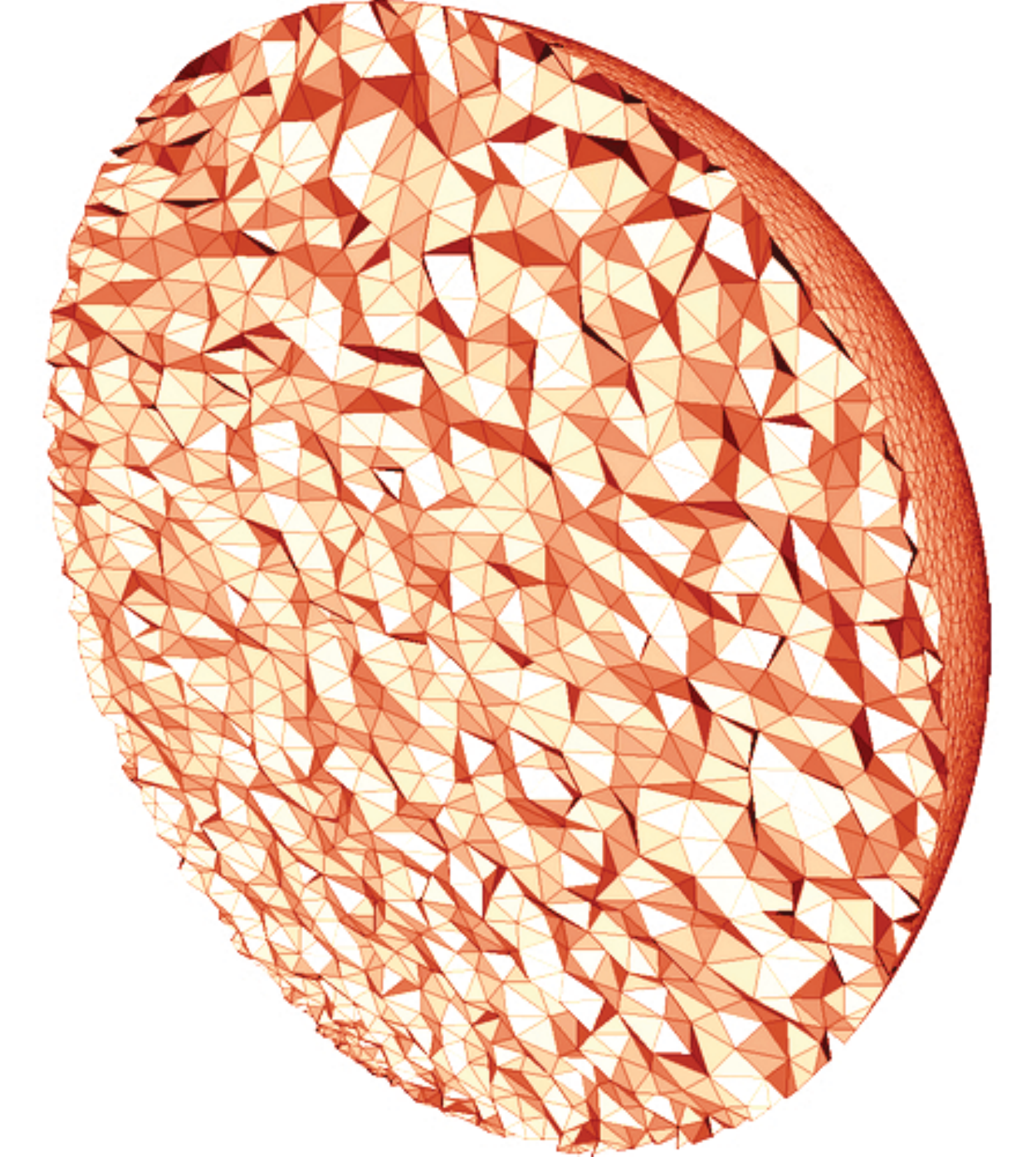}\\
 (a) & (b)
 \end{tabular}
 \caption{Examples of the tetrahedral mesh structure for (a) a prostate dataset and (b) the resulting ball from the volumetric curvature flow.  The mesh is cut through the center to show the interior structure.}
 \label{fig:tetMesh}
\end{figure}

The boundary mesh $\partial M$ has the initial induced Euclidean
metric $\mathbf{g}_0$ and the final metric  $\mathbf{g}_1$ on the
unit sphere. The discrete metrics are represented as edge lengths.
For each edge $e_{ij}$, we use $l_{ij}^0$ to denote its initial
length (metric), and $l_{ij}^1$ its final length on the sphere. Then
we define the metric at time $t$, $\mathbf{g}(t)$ as
\[
    l_{ij}^t = (1-t) l_{ij}^0 + t l_{ij}^1, \quad \forall [v_i,v_j]\in \partial M, \forall t\in
    [0,1].
\]

At each time $t$, we fix the boundary edge lengths and use Ricci
flow to adjust the interior edge lengths, such that all interior edge
curvatures become zeros. For each interior edge $e_{ij}$, we
deform its length as
\[
    \frac{d}{d\tau} l_{ij} = K_{ij}.
\]

After we compute the flat metric for the whole volume at time $t$, we
step further to compute that at time $t+\delta t$, using the
resulting metric at $t$ as the initial input, and boundary mesh
metric $\mathbf{g}(t)$ as the boundary condition. During the
evolvement, if we encounter a degenerated tetrahedron, we perform
local re-triangulation. See Algorithm~\ref{alg:a2} for details.

\begin{algorithm}[!]
\SetKwInOut{Require}{Require}
\SetKwInOut{Ensure}{Ensure}
\caption{{ Volume parameterization of a ball.}}
\Require{The prostate volume $M$, and boundary mapping $f:\partial M\to\mathbb{S}^2$}
\begin{enumerate}
\item Set $t=0$, the boundary metric to be the initial Euclidean\\
metric.
\item For all interior edges $e_{ij}$, compute the edge
curvature.
\item Evolve the interior edge length $l_{ij} = K_{ij} d\tau +
l_{ij}$.
\item If the tetrahedron is close to be degenerated, locally remesh \\
the neighborhood by swapping the shared face between two
adjacent tetrahedra.
\item Repeat step 2,3, until all the interior edge curvature are zeros.
\item Update the boundary metric. For each edge $e_{ij}$ on the
boundary, $l_{ij} = l_{ij} + (l_{ij}^1 - l_{ij}^0) \delta t$.
\item Update time $t=t+\delta t$.
\item Repeat step 2 through 5, until $t=1$.
\end{enumerate}
\normalsize
\label{alg:a2}
\end{algorithm}

Figure~\ref{fig:tetMesh}(b) shows the structure of the resulting tetrahedral mesh.  

\subsection{Volumetric Registration}
\label{Sec:reg}
After mapping different prostates to the unit solid ball, we register them using anatomical feature points.

\paragraph{Prostate Feature Detection}
The prostate, a gland like a walnut in size and shape, does not contain a complicated geometric structure. The prostate gland, which surrounds the urethra, is located in front of the rectum, and just below the bladder.

\begin{figure}[h]
\centering
\begin{tabular}{cc}
\includegraphics[height=0.22\textwidth]{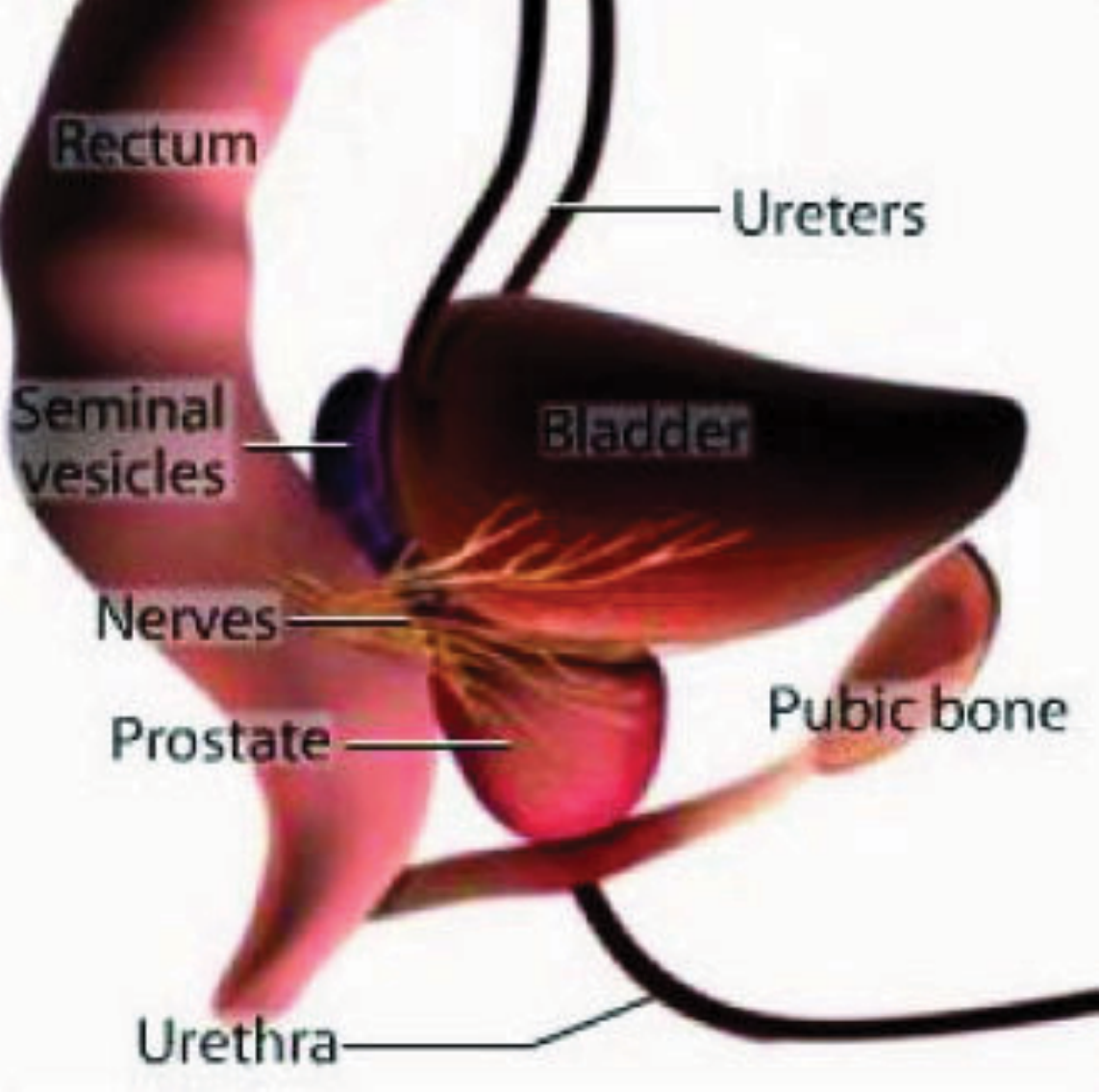} &
\includegraphics[height=0.22\textwidth]{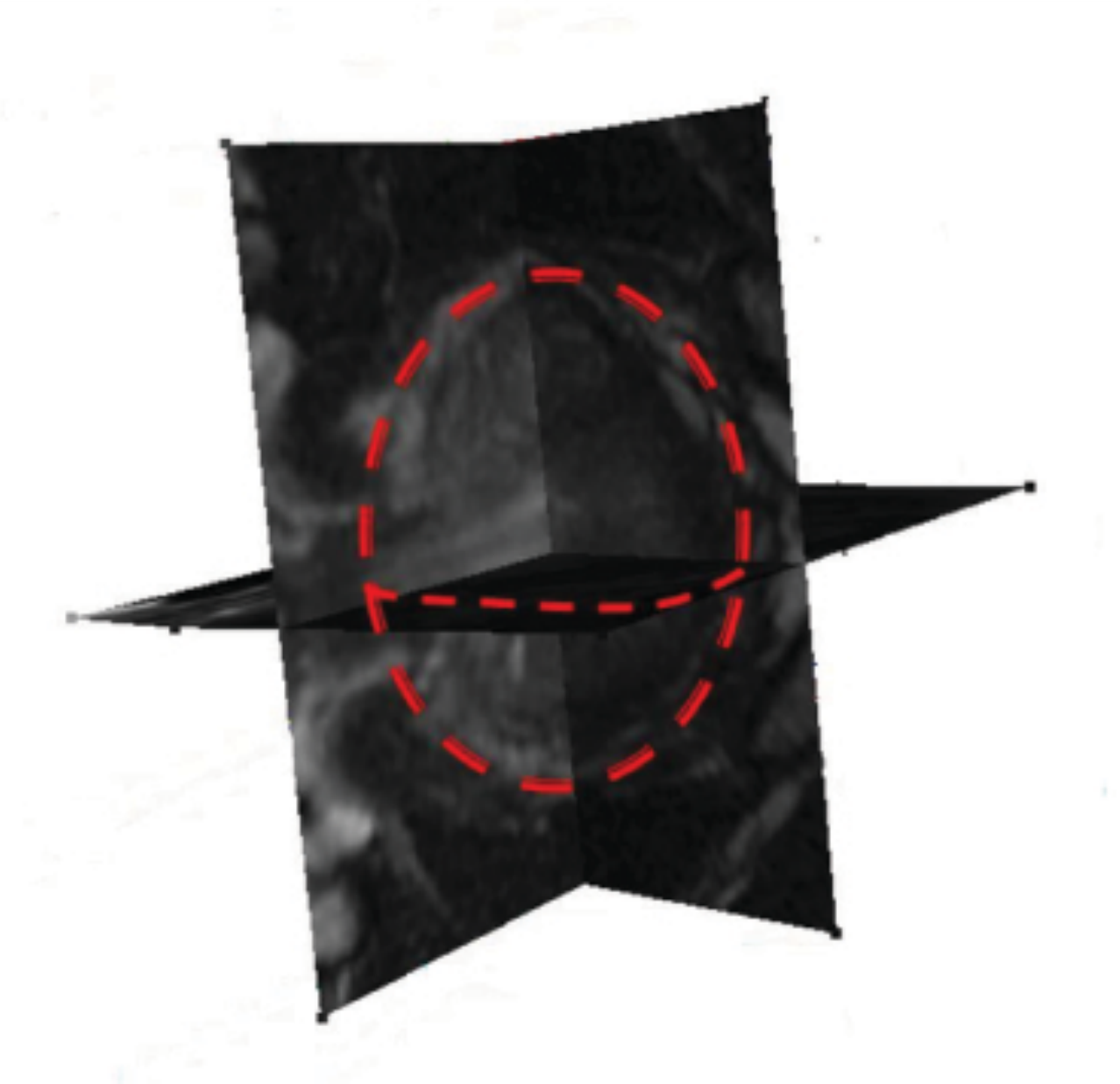} \\
(a) & (b) \\
\multicolumn{2}{c}{\includegraphics[height=0.30\textwidth]{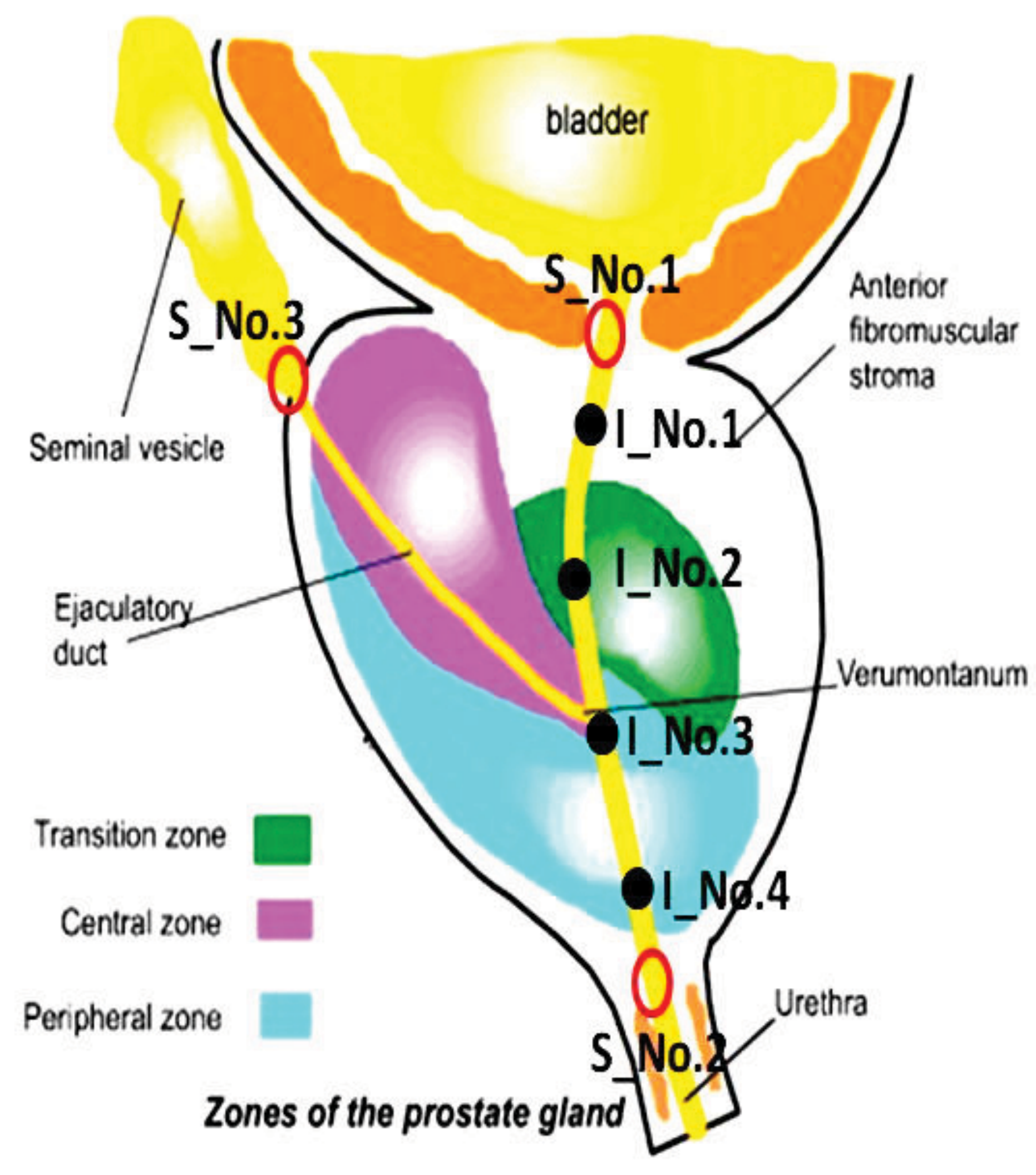}} \\
\multicolumn{2}{c}{(c)}
\end{tabular}
\caption{Anatomical prostate points. (a) The anatomical position of the prostate~\cite{rfig}. (b) Multi-view of a prostate MR image along three viewing directions. (c) The prostate structures~\cite{rfig}. Each feature point with the pre-defined index number is highlighted by the red circle. S\_No's are the surface feature points while \{I\_No\} is the interior feature point set. }
\label{fig:pfeatures}
\vspace{-12pt}
\end{figure}

For volumetric registration, we need to match at least three identical anatomical features within the MRI images of different directions to obtain the accurate and reliable registration result. A pair of glands called the seminal vesicles are tucked between the rectum and the bladder, and attached to the prostate as shown in Figure~\ref{fig:pfeatures}(a). The urethra goes through prostate and joins with two seminal vesicles at the ejaculatory ducts.
Therefore, some distinctive anatomical structures, such as the prostatic capsule and seminal vesicle contours, ejaculatory ducts, urethra, and dilated glands as represented in Figure~\ref{fig:pfeatures}(c), can be applied for the registration between different scan directions of one dataset or between MR slices and histology maps.

\begin{figure}[t]
\centering
\begin{tabular}{cc}
\includegraphics[height=0.22\textwidth]{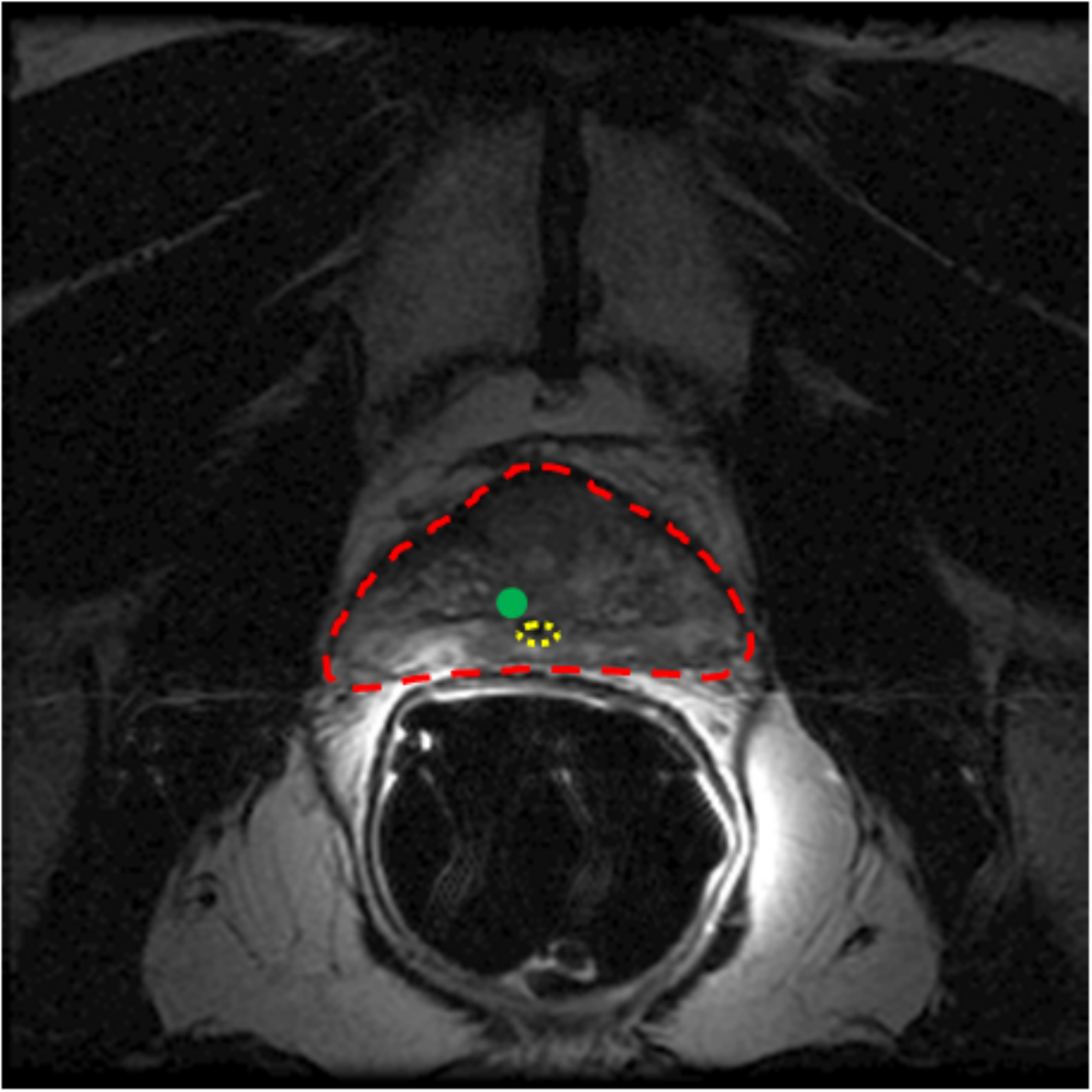} &
\includegraphics[height=0.22\textwidth]{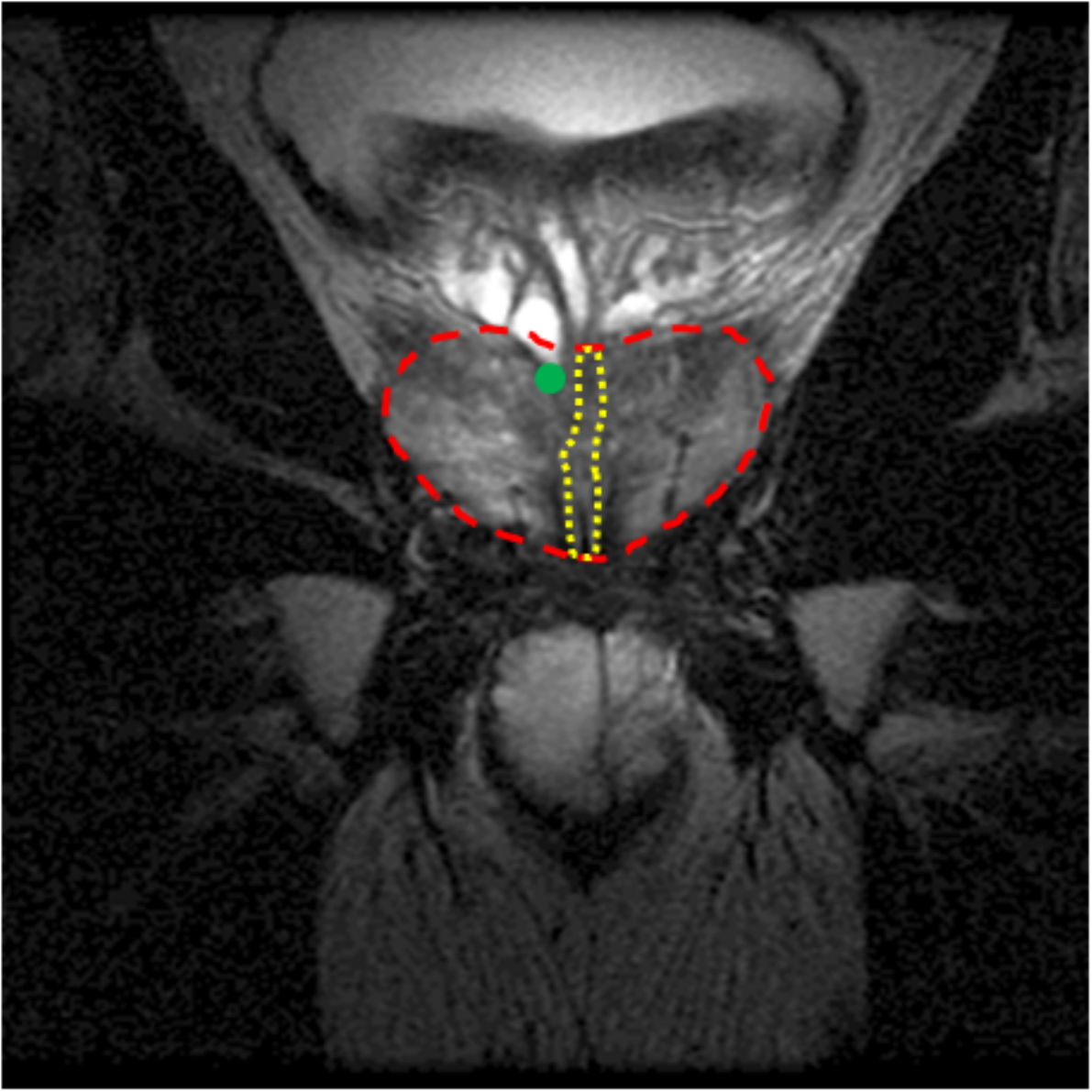} \\
(a) & (b) \\
\multicolumn{2}{c}{\includegraphics[height=0.28\textwidth]{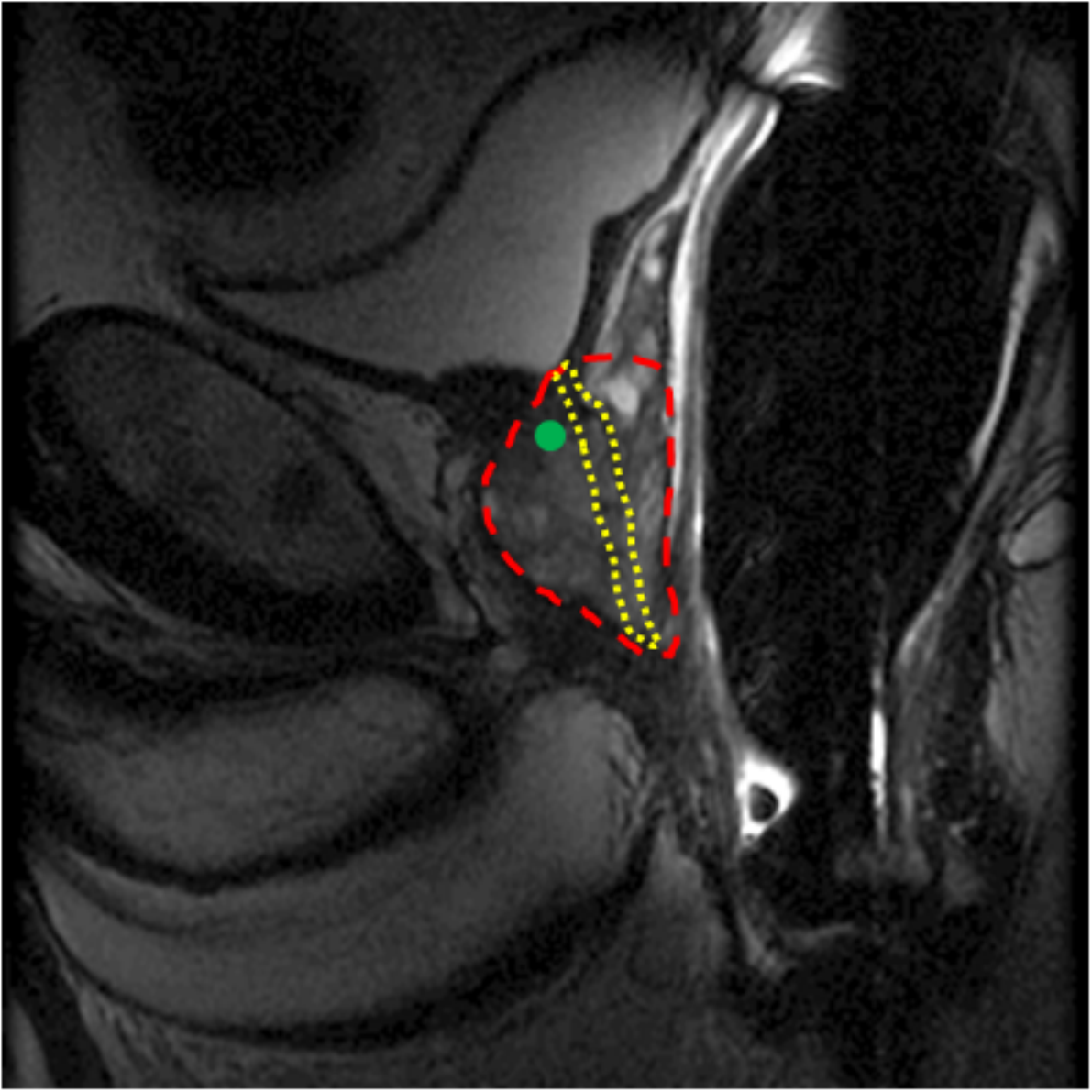}} \\
\multicolumn{2}{c}{(c)}
\end{tabular}
\caption{ Manually marked features on the MRI slices using (a) axial, (b) coronal and (c) sagittal view of the same prostate dataset. Red contour highlights the boundary of prostate and green points show the corresponding features. The urethra is marked using yellow contour.}
\label{fig:mriview}
\vspace{-12pt}
\end{figure}

MRI provides images with excellent anatomical
detail and soft-tissue contrast, and MRI sequences are displayed in a serial order. We analyze $T_{2}$-weighted datasets along the axial, sagittal and coronal view, as shown in Figure~\ref{fig:pfeatures}(b). On each MRI prostate view direction, the exact outline of prostate boundary is traced and all corresponding feature points are manually marked with predefined index numbers as shown in Figure~\ref{fig:mriview}.
For registration, we use three exterior feature points and a set of interior ones based on the structure information of urethra and seminal vesicles: two surface feature points are exactly the entrance and exit points of the urethra because the urethra goes through the entire prostate, while the third surface point is marked at the intersection between each seminal vesicle and prostate with respect to the fact that two seminal vesicles attach to the surface of prostate and merge with urethra at the ejaculatory ducts. A set of interior feature points are marked along the urethra, beginning with the entrance surface feature point and ending with the exit one.

\paragraph{Registration Framework}
Let $M_1$ and $M_2$ denote the volumes of the two prostates. The
computational algorithm for registration is as shown in the following diagram.
\[
\begin{CD}
M_1 @>\phi >> M_2\\
@VV \phi_1 V @VV \phi_2V\\
\mathbb{D}^3 @>\eta >> \mathbb{D}^3
\end{CD}
\]
We first compute two volumetric maps $\phi_1: M_1 \to \mathbb{D}^3$,
and $\phi_2: M_2 \to \mathbb{D}^3$. Then, we compute an automorphism
of $\mathbb{D}^3$, $\eta: \mathbb{D}^3\to \mathbb{D}^3$, which
aligns all the feature points. Then the final registration is given
by
\[
    \phi = \phi_2^{-1}\circ\eta\circ\phi_1.
\]
The key component is to construct $\eta$. Suppose $\{p_1,p_2,\cdots,
p_n\}$ and $\{q_1,q_2,\cdots, q_n\}$ are the corresponding feature
points on $\phi_1(M_1)$ and $\phi_2(M_2)$. Then $\eta$ should align
them $\eta(p_k)=q_k$.

\begin{figure*}[ht!]
\centering
\begin{tabular}{c}
\includegraphics[width=6in]{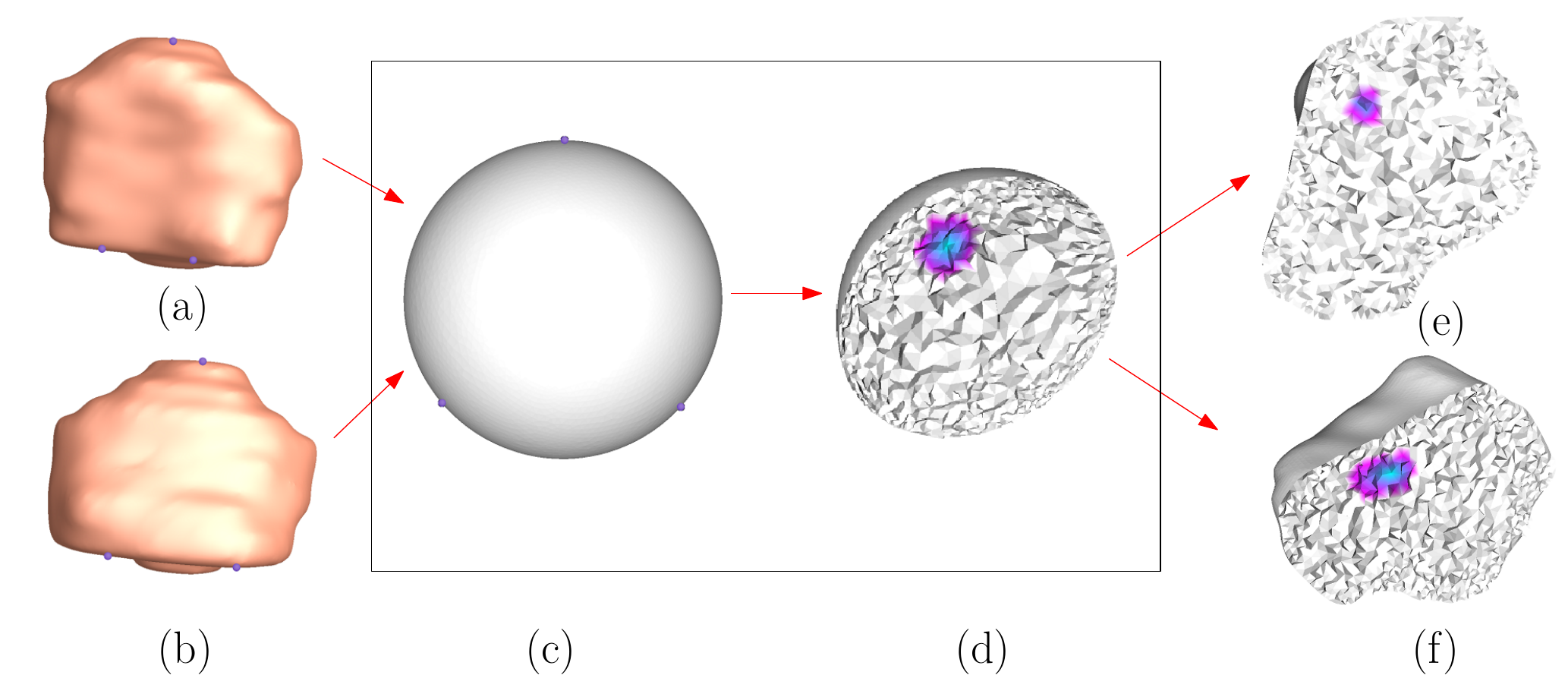}
\end{tabular}
\caption{Registration of $T_2$ and $T_2$ coronal view MR scans for the same patient at (a) day 102 and (b) day 1801. MR scans are taken in $T_2$ coronal modes and segmented. These segments are then reconstructed into a surface and mapped onto respective spheres and tetrahedralized. These spheres are then registered using surface feature points on another sphere (c). On this registered sphere, we mark regions (d) and since our mapping is diffeomorphic, these regions are mapped back to the original tetrahedralized prostate shapes (e \& f).}
\label{fig:T2-T2-reg}
\end{figure*}

First we choose three feature points on the boundaries (see e.g. Figure~\ref{fig:T2-T2-reg}(a)). Assume
$\{p_1,p_2,p_3\}$ and $\{q_1,q_2,q_3\}$ are the three points on the two volumes. We use stereo-graphic
projection $\pi:\mathbb{S}^2\to \mathbb{C}$ to map the boundary of
the ball onto the complex plane. Then, we construct a M\"obius
transformation to align these three feature points. By abusing the
symbols, we still use $p_k,q_i$ for their images of the
stereo-graphic projection on the complex plane. Define
\[
    \rho_1(z) = \frac{(z-p_1)(p_3-p_2)}{(z-p_2)(p_3-p1)}
\]
then $\rho_1$ maps $p_1,p_2,p_3$ to $0,\infty,1$. Similarly, we
define $\rho_2$ which maps $q_1,q_2,q_3$ to $0,\infty,1$. Then
$\rho_2^{-1}\circ \rho_1$ aligns there feature points on the complex
plane, $ \pi^{-1}\circ \rho_2^{-1}\circ \rho_1\circ \pi$ align the
feature points on the sphere.

In order to align interior feature points, we add position and target curvature constraints to volumetric curvature flow. Then in the parameter ball, the interior feature points will be placed at exactly the target position. To align the feature points, we first map model A to a solid sphere with no constrains, and get the result coordinates of the features; then map model B to the solid sphere with target feature position the same as these feature coordinates in the solid sphere. By doing this, we can perfectly align these feature points, which leads to a good alignment of the models. Furthermore, our method does not create flipped tetrahedrons as in the case of harmonic mapping (as explained in the following section), so the registration result is good, both locally and globally.

\section{Evaluation}
\label{Sec:eval}

\begin{figure}[tb]
    \centering
 \begin{tabular}{cc}
\includegraphics[height=0.22\textwidth]{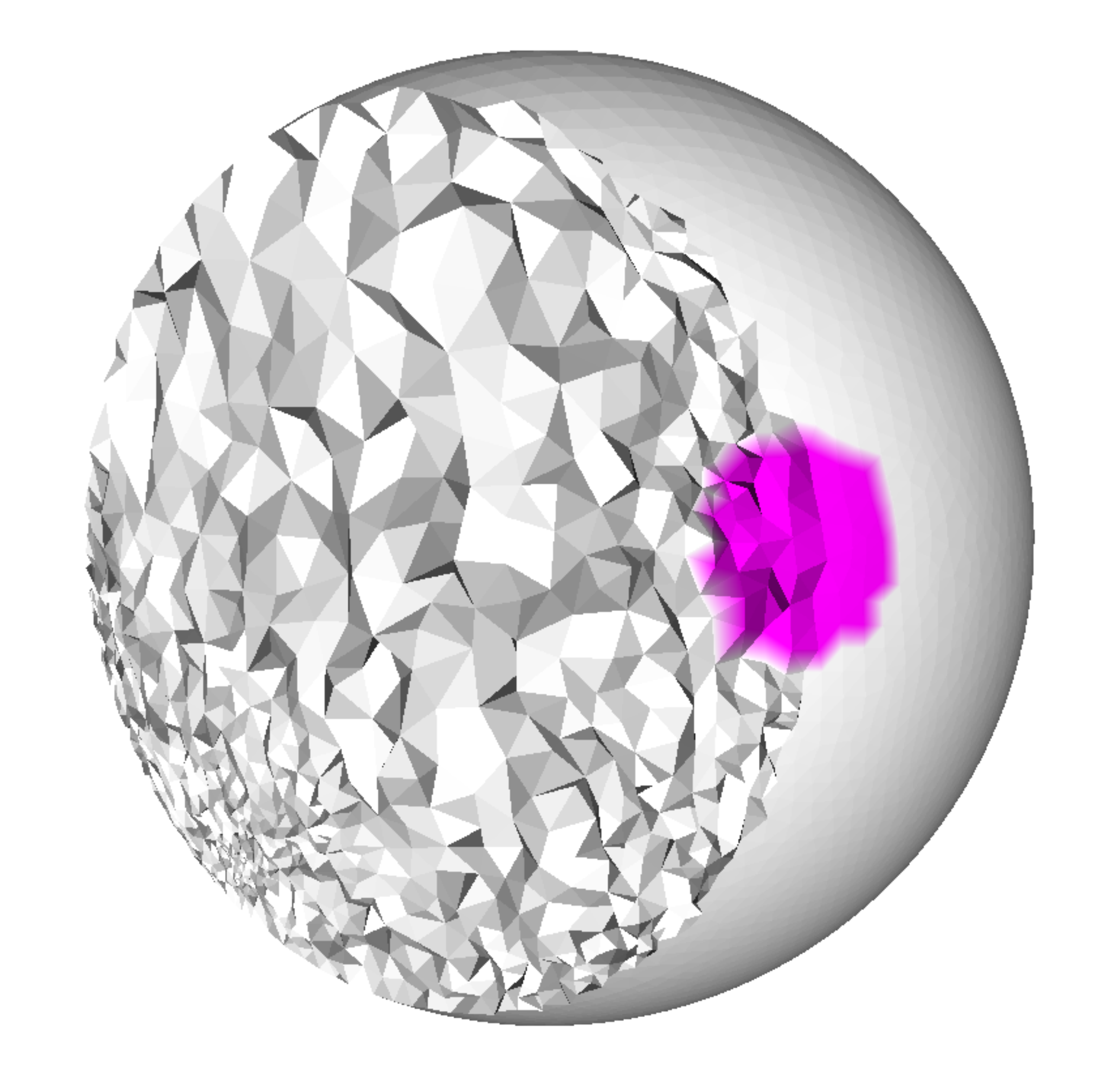} &
\includegraphics[height=0.22\textwidth]{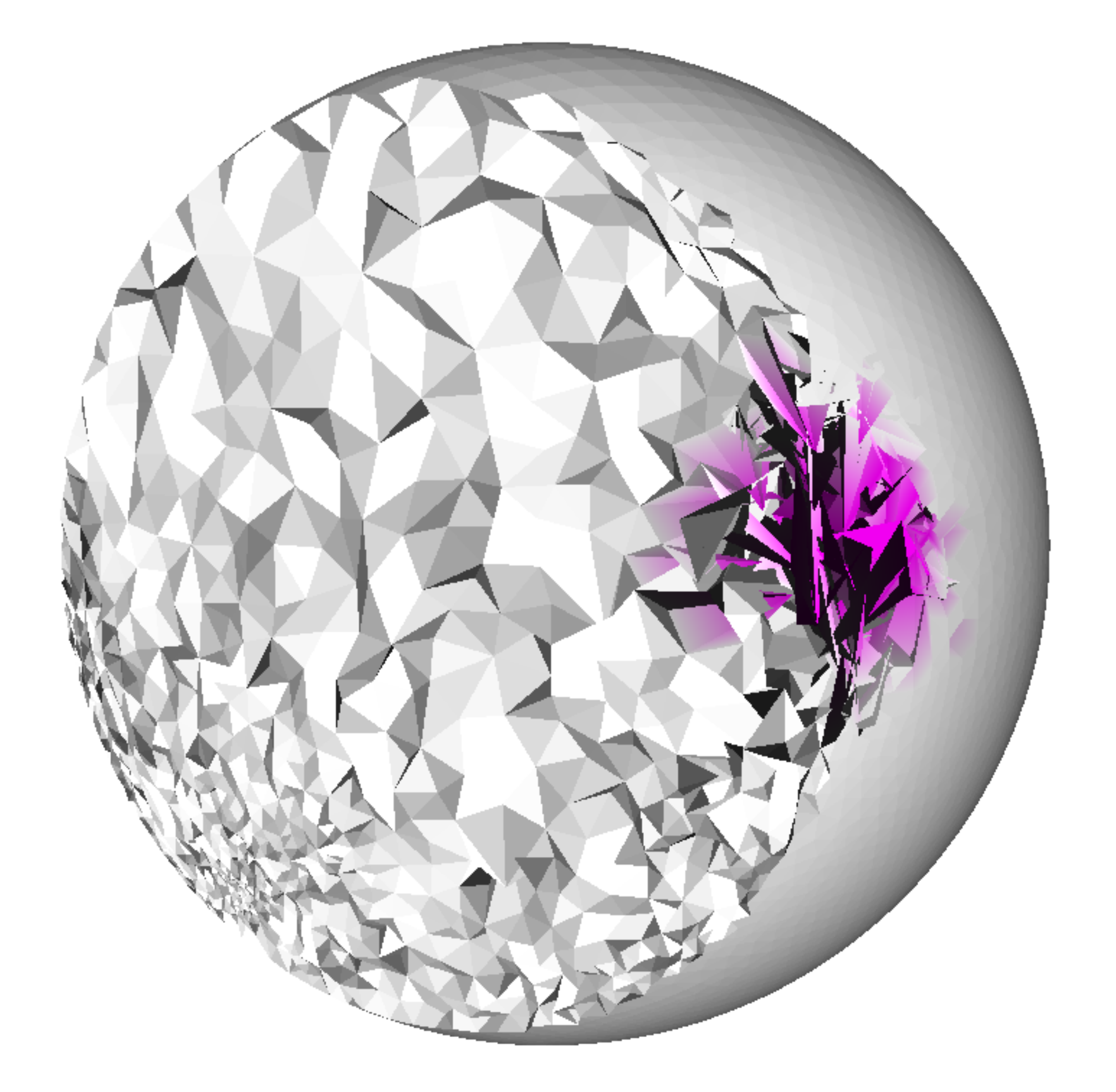} \\
 (a) & (b)
 \end{tabular}
 \caption{The comparison of (a) our ricci flow method with (b) harmonic mapping. The edges are flipped at the boundary in harmonic mapping, making in unsuitable for our application.}
 \label{fig:comparison}
\end{figure}

We test the algorithm using $T_1$ and $T_2$-weighted MR postate images. In $T_1$ it is not possible to identify interior feature points, so wherever we consider $T_1$ for registration, we make use of only the surface feature points. In $T_2$, we can identify interior feature points so we try to identify as many of these as possible on the corresponding prostates during the registration. We also observe that the urethra is easily identified in the coronal view (in high quality MR Images) so to take advantage of the feature points along urethra, whenever possible, we use $T_2$ coronal views. We also compare our method against harmonic maps.

Harmonic map method has been broadly applied for surface and volume registration in medical imaging. One of the major merits for surface harmonic map is that it produces diffeomorphism, as stated in Rado's theorem \cite{SchoenLectures:1997}: for a harmonic map from a surface with the disk topology to a convex planar domain, if the restriction of the map on the boundary is a homeomorphism, then the interior harmonic map is a diffeomorphism. Unfortunately, this theorem doesn't hold for volumetric case. As shown in Fig. \ref{fig:comparison}, the volumetric harmonic map introduces flipped tetrehedra (the orientation of the image tetrhedra is reversed) near the boundary area. In contrast, our current method guarantees the mapping to be one-to-one and flipping-free.

\begin{figure*}[ht!]
\centering
\begin{tabular}{c}
\includegraphics[width=4.5in]{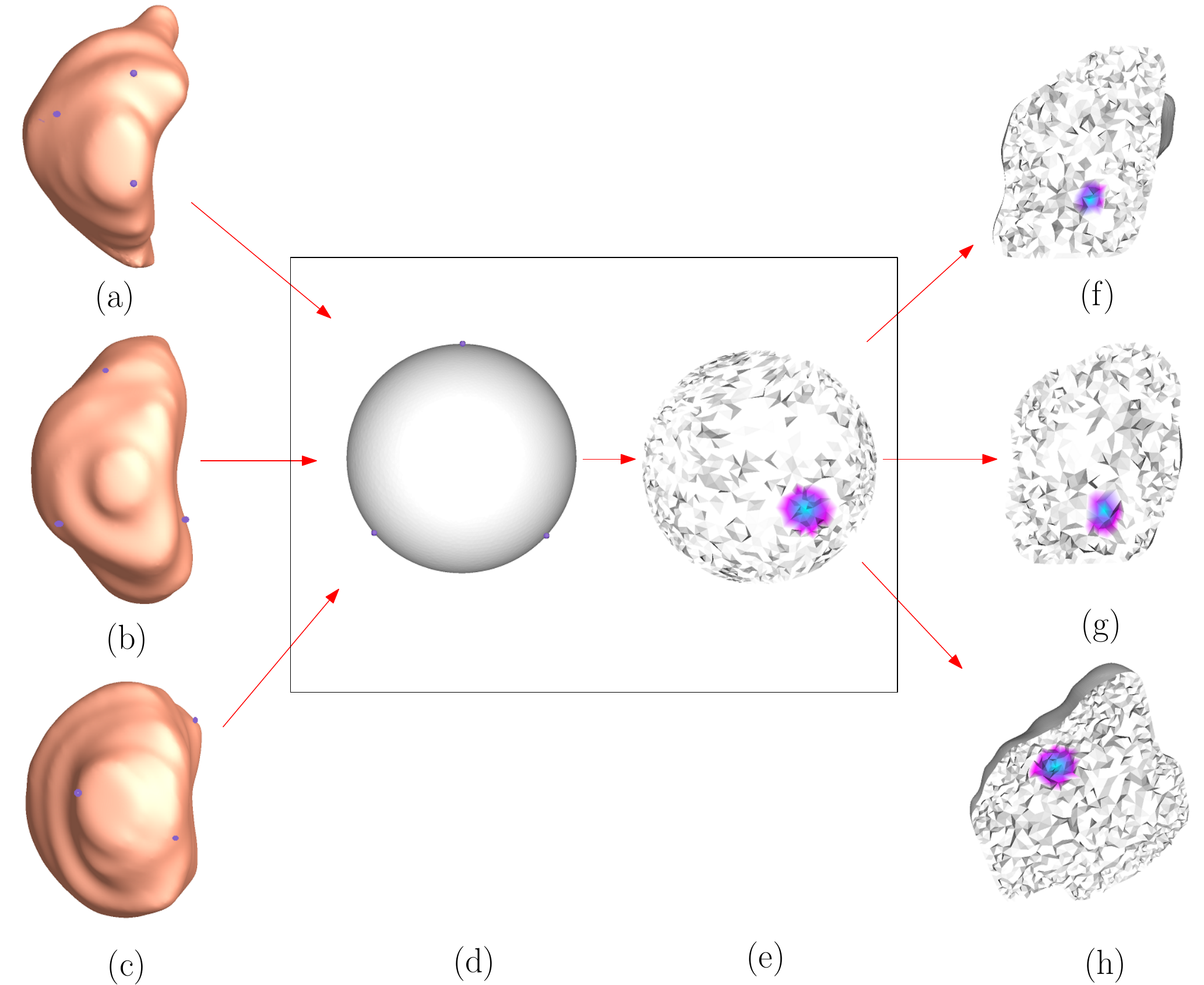}
\end{tabular}
\caption{Registration of $T_2$-weighted MR scans of the same patient in different orientations. (a) Axial view, (b) Coronal view and (c) Sagittal view. These are registered and tetrahedralized (d). A region is marked on the registered sphere (e) and the corresponding regions on the registered prostate models are identified (f,g \& h).}
\label{fig:orientations}
\end{figure*} 
\section{Results}

The data we have used for this work is from ACRIN 6659 study, the Prostate MR Image Database \cite{MRIDatabase:2008} and Stony Brook UHMC Database. Our work focuses on the registration, so for this project we have used manual segmentation (done by an expert/radiologist) to obtain the prostate gland. We will make the data from Stony Brook UHMC Database publicly available as well.

The Prostate MR Image Database contains MR prostate scans for 231 patients with multiple days and in different modes. The Prostate MR Image Database also contains expert segmentations for most of the datasets, but there were instances where we were not able to find the segmentation corresponding to a specific dataset. For these instances, we employed the help of a radiologist to do the segmentation. The Stony Brook UHMC Database contains MR scans for more than 200 patients. In this case, there were no prior segmentations available so a radiologist segmented out the datasets tested.

\begin{table}[h]
	\centering
  \caption{Running time of the volumetric curvature flow algorithm for the three prostate MR scans, corresponding to the models reconstructed in Figure \ref{fig:orientations}(a), (b) \& (c).}
  \scalebox{0.85}{
  \begin{tabular}{|c|c|c|c|c|c|}
  \hline
  MRI scans & Vertices & Tetrahedra & Edges & Faces & Running time \\
  \hline
  a & 96536  & 564478 &678062 & 1146005 & 170 secs \\
  b & 98898  & 580013  & 695976 &1177092 & 191 secs \\
  c & 115946   & 689982   & 822853  &1396890  & 210 secs \\
	\hline
  \end{tabular}}
  \label{table:meshSizes}
\end{table}

This work was performed on an Intel Xeon E2540 2.5GHz machine with 16GB of RAM. The running time to map the mesh from its original shape to the sphere is around 3 minutes for the 3 MRI scans for Figure \ref{fig:orientations} (see Table~\ref{table:meshSizes}).

\begin{table}[h]
	\centering
  \caption{Registration error computed with respect to the Euclidean distances between the registered feature points on different prostates (Figure \ref{fig:urethra}), then divided by the diagonal length of the model's bounding box (in this case, the length is 100). The Feature\# are the labels shown in Figure \ref{fig:urethra}(a). The second column shows the error distances computed between feature points in Figure \ref{fig:urethra}(a) and (b), whereas the third column shows the distances between Figure \ref{fig:urethra}(a) and (c).}
  \label{table:reg-error}
  \begin{tabular}{|c|c|c|}
  \hline
  Feature\# & Reg Error & Reg Error \\
  && w/ seg distortion \\
  \hline
    1 & 0.0035 & 0.0119 \\
    2 & 0.0111 & 0.0148 \\
    3 & 0.0096 & 0.0108 \\
    4 & 0.0089 & 0.0091 \\
    5 & 0.0053 & 0.0078 \\
    6 & 0.0032 & 0.0051 \\
    7 & 0.0196 & 0.0201 \\
    8 & 0.0187 & 0.0221 \\
    9 & 0.0117 & 0.0175 \\
    10 & 0.0112 & 0.0129 \\
    11 & 0.0148 & 0.0174 \\
    12 & 0.0106 & 0.0181 \\
    13 & 0.0179 & 0.0209 \\
  \hline
  \end{tabular}
\end{table}

\textbf{\emph{Different Days}}. We have tested our method with different modes of MRI, i.e. $T_1$ and $T_2$-weighted scans. For prognosis monitoring of a tumor, we register $T_2$ coronal views from multiple days (as shown in Figure \ref{fig:T2-T2-reg}) since the urethra is easily visible in this view, as discussed in Section \ref{Sec:eval} above. Then we can identify the reduced intensity regions, if found, and mark the corresponding region on the registered sphere. Since our mapping is diffeomorphic, we can get corresponding regions on the registered prostate models, as shown in Figure \ref{fig:T2-T2-reg}. We see that the suspected region is larger at Day 1801 than at Day 102. This allows us to measure the shape of the tumor, if marked accurately. Furthermore, we can register as many scans as possible, preferably $T_2$-weighted in coronal view, to get a continuous progression of the abnormality. In case of Figure \ref{fig:T2-T2-reg}, we have registered these prostate models on the basis of four feature points, 3 surface feature points and one interior feature point (marked at the intersection of three lines of urethra). We then mark out the other feature points along urethra to study the error.

\begin{figure}[tb]
    \centering
 \includegraphics[width=3in]{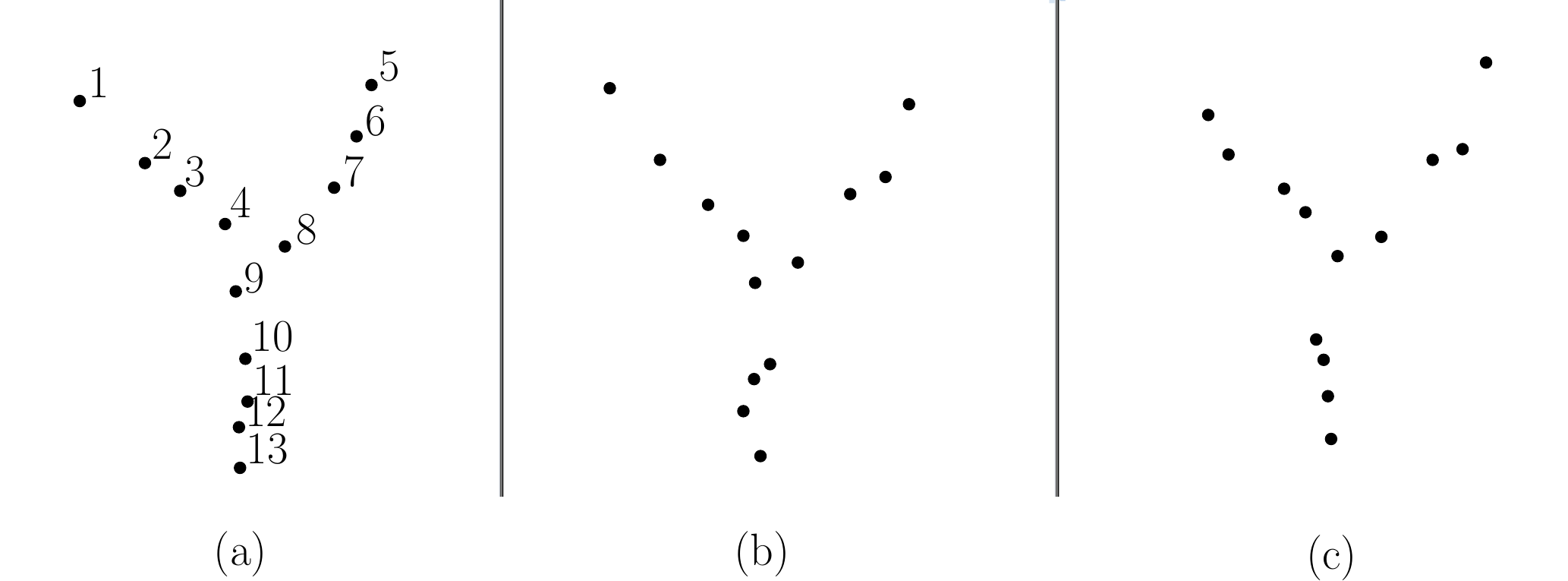}
 \caption{(a) The feature points marked out for the urethra on the prostate after registration in Figure \ref{fig:T2-T2-reg}(a). (b) The corresponding urethra feature points on the prostate (b) in Figure \ref{fig:T2-T2-reg}(b) after registration. (c) The corresponding urethra feature points after a segment was removed from the original MR scans.}
 \label{fig:urethra}
\end{figure}

For the alignment of these feature points along urethra, we actually create a one-to-one map between prostate models A (Figure \ref{fig:T2-T2-reg}(a)) and B (Figure \ref{fig:T2-T2-reg}(a)) by mapping these two models to the same parameter sphere with the original 4 features aligned. Hence, we map the urethra feature points in model B to model A by first mapping them to the parameter sphere and then to model A. As shown in Figure \ref{fig:urethra}, the original urethra feature points in model A are well aligned with that from model B, which gives a strong evidence of our good registration quality. We further provide the evidence of the quality of our registration using the euclidean distances computed between these points as shown in Table \ref{table:reg-error}. Since we are using manual segmentations, we also provide the error if the segment boundary is distorted and a complete segment slice is removed. The results for this combined distortion and removal of a segment slice are shown as the $3rd$ column of Table \ref{table:reg-error} as well. As seen, the Euclidean distances does not change drastically giving the strong foundations for our registration.

\textbf{\emph{Different Modes of MRI}}. Moreover, we also register different modes of MR scans for a patient on the same day. The result of this registration is shown in Figure \ref{fig:teaser}. We register $T_1$ and $T_2$ scans in axial view on the basis of 3 surface feature points, since interior feature points can not be identified in $T_1$. We observe that if the interior feature points are marked on the registered sphere from $T_2$ scans, we can get a good estimate for the location of these interior feature points on the registered prostate model, reconstructed from $T_1$ scans (given the good quality of our registration regardless of the minor distortions in the initial segmentation). To prove this is the case, we registered 3 $T_2$-weighted MR scans in axial view from the same day for a patient on the basis of just the surface feature points and then computed the euclidean distances (as done above) to study the error if a feature point is marked at the intersection of the three lines of urethra on one of the prostate models and the rest are registered with reference to this. The error computed did not exceed $0.00360$ in this case (again the diagonal length of the bounding box was 100 as in Figure \ref{fig:urethra}), showing strong evidence that given good segmentation, the feature points from one prostate model can be accurately marked out on the other registered models.

\textbf{\emph{Different Orientations}}. We also register MR scans acquired in different orientations, as shown in Figure \ref{fig:orientations}. The axial, coronal, and sagittal images are all segmented, reconstructed, and parameterized. The resulting models are then registered using the feature points stated in Section~\ref{Sec:reg}. Table~\ref{table:reg-error1} shows the error registration for three extra geometric identical points used to evaluate the quality of registration. We use similar method as discussed above for error evaluation by keeping one model as a reference and mapping the identical points from the other to the reference and compute the euclidean distances.

From the results, we observe that the error for feature points in sagittal and coronal views is less than the axial view and the other views. This is due to the error of segmentation and feature points location. In the coronal or sagittal direction, all the feature points are within one or two slices, and we are able to precisely distinguish their locations. However, in the axial direction, features are scattered in multiple slices. Since the interslice resolution is much lower than the intraslice one, the euclidean distance error for the extra geometrical feature points is larger in the axial and sagittal/coronal combination than the coronal and sagittal combination. However, still, our results show good quality of our registration even in this scenario.

\begin{table}[h]
	\centering
  \caption{Registration error computed with respect to the Euclidean distances between the registered feature points on different prostates (Figure \ref{fig:orientations}), then divided by the diagonal length of the model's bounding box (in this case, the length is 100 again). The feature\# are the labels of the identical geometric feature points to evaluate the method. The second column shows the error distances computed between feature points in Figure \ref{fig:orientations}(a) Axial and (b) Coronal, the third column shows the distances between Figure \ref{fig:orientations}(a) Axial and (c) Coronal and the fourth column shows the error between feature points in Figure \ref{fig:orientations}(b) Coronal and (c) Sagittal}
  \begin{tabular}{|c|c|c|c|}
  \hline
  Feature\# & Reg Error & Reg Error & Reg Error \\
  & (a) \& (b) & (a) \& (c) & (b) \& (c) \\
  \hline
    1 & 0.0025 & 0.0029 & 0.0014 \\
    2 & 0.0021 & 0.0024 & 0.0018 \\
    3 & 0.0022 & 0.0027 & 0.0019 \\
  \hline
  \end{tabular}
  \label{table:reg-error1}
\end{table} 
\section{Conclusion}

In this work, we performed a registration of volumetric meshes of the prostate gland using volumetric curvature flow. We have tested our method by registering the MR scans of the same patient in different orientations, from different days and with different modes of MRI. Our method gives good results on data from the ACRIN 6659 study, Prostate MR Image Database and Stony Brook UHMC Database. Moreover, analysis of the resulting registration shows accurate demarcation of the same region can be achieved across different registered prostates (from different scans), even with distortions in the initial segmentation.

In the future, we plan to integrate our registration method with automatic/semi-automatic segmentation to create a complete framework for the prostate and eventually, extend this framework for other volumetric objects, such as the brain. We also plan to validate our method across different modalities, such as CT, MRI and ultrasound. 

\acknowledgments{
The data we have use for this work is from ACRIN 6659 study, the Prostate MR Image Database \cite{MRIDatabase:2008} and Stony Brook UHMC Database. This work has been supported by NSF grants IIS0916235, CCF-0702699 and CNS0959979 and NIH grant R01EB7530.}

\end{document}